\documentclass[prx, aps, nofootinbib, twocolumn, amssymb, superscriptaddress, 10pt]{revtex4-2}
\usepackage{amsmath}
\usepackage{amssymb}
\usepackage{amsthm}
\usepackage{amsfonts}
\usepackage{enumerate}
\usepackage{latexsym}
\usepackage{color}
\usepackage{setspace} 
\usepackage{blindtext}
\usepackage{dsfont}
\usepackage{mathrsfs}
\usepackage[normalem]{ulem}

\usepackage{tabularx}
\newcolumntype{Y}{>{\centering\arraybackslash}X}

\usepackage{stackengine}

\usepackage{bm}
\usepackage{graphicx}

\usepackage{hyperref}
\hypersetup{
pdfnewwindow=true, colorlinks=true,
linkcolor=amethyst, anchorcolor=amethyst,
citecolor=amethyst, filecolor=amethyst,
menucolor=amethyst, urlcolor=amethyst} 

\usepackage{pifont}

\newcommand{\beginsupplement}{
        \setcounter{table}{0}
        \renewcommand{\thetable}{S\arabic{table}}
        \setcounter{figure}{0}
        \renewcommand{\thefigure}{S\arabic{figure}}
        \setcounter{equation}{0}
        \renewcommand{\theequation}{S\arabic{equation}}
        \setcounter{section}{0}
        \renewcommand{\thesection}{\Alph{section}}
        \setcounter{subsection}{0}
        \renewcommand{\thesubsection}{\arabic{subsection}}
}

\newcommand{\vk}{{\mathbf{k}}}

\newcommand{\vq}{{\mathbf{q}}}

\definecolor{amethyst}{rgb}{0.6, 0.4, 0.8}

\begin{document}

\title{
Chern Bands' Optimally Localized Wannier Functions and Fractional Chern Insulators
}
\author{Fang Xie}
\email{fx7@rice.edu}
\affiliation{Department of Physics \& Astronomy, Rice University, Houston, TX 77005, USA}
\affiliation{Rice Academy of Fellows, Rice University, Houston, TX 77005, USA}
\author{Yuan Fang}
\affiliation{Department of Physics \& Astronomy, Rice University, Houston, TX 77005, USA}
\author{Lei Chen}
\affiliation{Department of Physics \& Astronomy, Rice University, Houston, TX 77005, USA}
\author{Jennifer Cano}
\affiliation{Department of Physics and Astronomy, Stony Brook University, Stony Brook, NY 11794, USA}
\affiliation{Center for Computational Quantum Physics, Flatiron Institute, New York, NY 10010, USA}
\author{Qimiao Si}
\affiliation{Department of Physics \& Astronomy, Rice University, Houston, TX 77005, USA}

\date{\today}

\begin{abstract}
Recent development on fractional Chern insulators and proximate phases call for a real space representation of isolated Chern bands. Here we propose a new method for a general construction of optimally localized Wannier functions from such Chern bands. We do so through an optimal gauge choice of the Bloch states of a Chern band with the singularity placed at any desired position in momentum space.
We apply this method to construct the optimally localized Wannier functions for kagome lattice, and use it to identify channels of interactions that are favorable to the development of fractional Chern insulators.
Implications of the approach for the interplay between correlations and topology in broader contexts are discussed.
\end{abstract}

\maketitle

{\it Introduction.} 
Correlated flat bands provide a rich setting to realize a variety of quantum phases and their transitions \cite{Stormer-rmp99,KirchnerRMP2020,Bistritzer2011Moire,cao_correlated_2018}.
Recent years have seen increasing recognition of the 
potential of the correlated flat band materials for merging the topology of electronic wavefunctions and strong correlations that go beyond the single particle picture
\cite{Pas21.1,Checkelsky-NRM2024}.
In addition to tunable flat bands of moir\'{e} systems \cite{Bistritzer2011Moire,cao_correlated_2018}, 
there has been rapid recent progress on flat band physics in frustrated lattice systems with electron motions experiencing a destructive interference \cite{Huang-Natphys2024,Ye-Natphys2024,Ekahana2024Anomalous}.
In systems such as kagome (and pyrochlore) metals, isolated topological flat bands develop from the interplay between lattice geometry and spin-orbit coupling. When the Coulomb interaction is large enough to mix the flat and accompanying wide bands (but still smaller than the width of the wide bands), the notion of compact molecular orbitals allows for the representation of the flat and wide bands in real space \cite{chen2023Metallic,Chen_emergent_2024,HuSciAdv2023}.
The molecular orbitals, exponentially localized, provide a means to treat the effect of Coulomb interactions in terms of a topological Kondo lattice \cite{lai2018weyl,chen2022topological}. 
This route has led to a phase diagram containing a quantum critical point with strange metallicity. The theoretical prediction has since been supported by pressure-tuning experiments in a new kagome metal
\cite{Liu-CsCr3Sb5-2023}.

An alternative regime is realized when the Coulomb interaction is larger than the flat bandwidth but still smaller than both the energy gap separating the flat and wide bands and the width of the wide bands. 
In this regime, the flat bands are to be considered in isolation.
For kagome metals, it is this regime which pertains to fractional Chern insulators (FCI) \cite{Tang2011High}.

To be specific, consider the case of a flat Chern band, with width $D_{\rm flat}$, the energy gap $\Delta_{g}$ that separates it from wide bands, and the interaction strength $U$.
When these three energy scales satisfy the following hierarchy, the system can be described by keeping only the Chern band states as the active Hilbert 
space
\cite{Setty2021Electron,Lin2023Complex}:
\begin{equation}\label{eqn:hierarchy}
    (D_{\rm flat}\,, U) \ll \Delta_g\,.
\end{equation}
To treat the effect of interactions, it is desirable to represent the bands in real space.
The projection of the Coulomb interactions in the real space basis is important not only for studying the 
non-perturbative effects of the interactions in general, but also to analyze the competition between the fractional Chern insulator and other correlated phases.

{\it General construction of Wannier functions.}
Wannier functions, which are defined as the discrete Fourier transformations of Bloch states \cite{ashcroft1976solid}, form an orthogonal complete basis in the Hilbert space spanned by the bands of interest.
When the gauge choice of the Bloch states are smooth over the whole Brillouin zone, the Wannier functions can be exponentially localized in real space \cite{rudin1973functional}. As such, the Wannier functions allow for the construction of a real space effective model \cite{marzari_maximally_2012}. 
In one dimensional systems, one can always construct a smooth gauge for a band that is gapped from other bands by finding the eigenstates of the projected position operator to minimize the spread of the Wannier function \cite{Marzari1997Maximally}.
In two dimensions, in contrast, such a construction fails due to the noncommutativity of the two components of the projected position operator.
Thus, numerical methods are usually necessary to find optimal Wannier functions in two dimensions. 
Another closely related issue is the difficulty in using Wannier functions when there is topological obstruction. 
To overcome the difficulty, one needs to consider the complementary bands to trivialize the band topology, as discussed for example in the context of twisted bilayer graphene \cite{Po2018Origin,Po2019Faithful,Carr2019Derivation}.
When the topological nontrivial band is considered alone, the existence of a non-zero Chern number in a two dimensional system prohibits a globally smooth gauge, and the Wannier function of a Chern band can only exhibit a power-law decay in real space \cite{Thouless1982Quantized, thouless_wannier_1984, Monaco2016Localization, Monaco2018Optimal}. 
In using such Wannier functions, minimizing the spread becomes extremely important. 
For certain Chern band systems, including the lowest Landau levels and twisted bilayer graphene, the construction of their Wannier functions have been discussed in literature \cite{Rashba1997Orthogonal, Panfilov2016chiral, zang_real_2022}.
However, a general construction of the optimal Wannier function in a generic Chern band, which also corresponds to ``minimizing'' its spread, remains an under-explored field.

\begin{figure}[t]
    \centering
    \includegraphics[width=\linewidth]{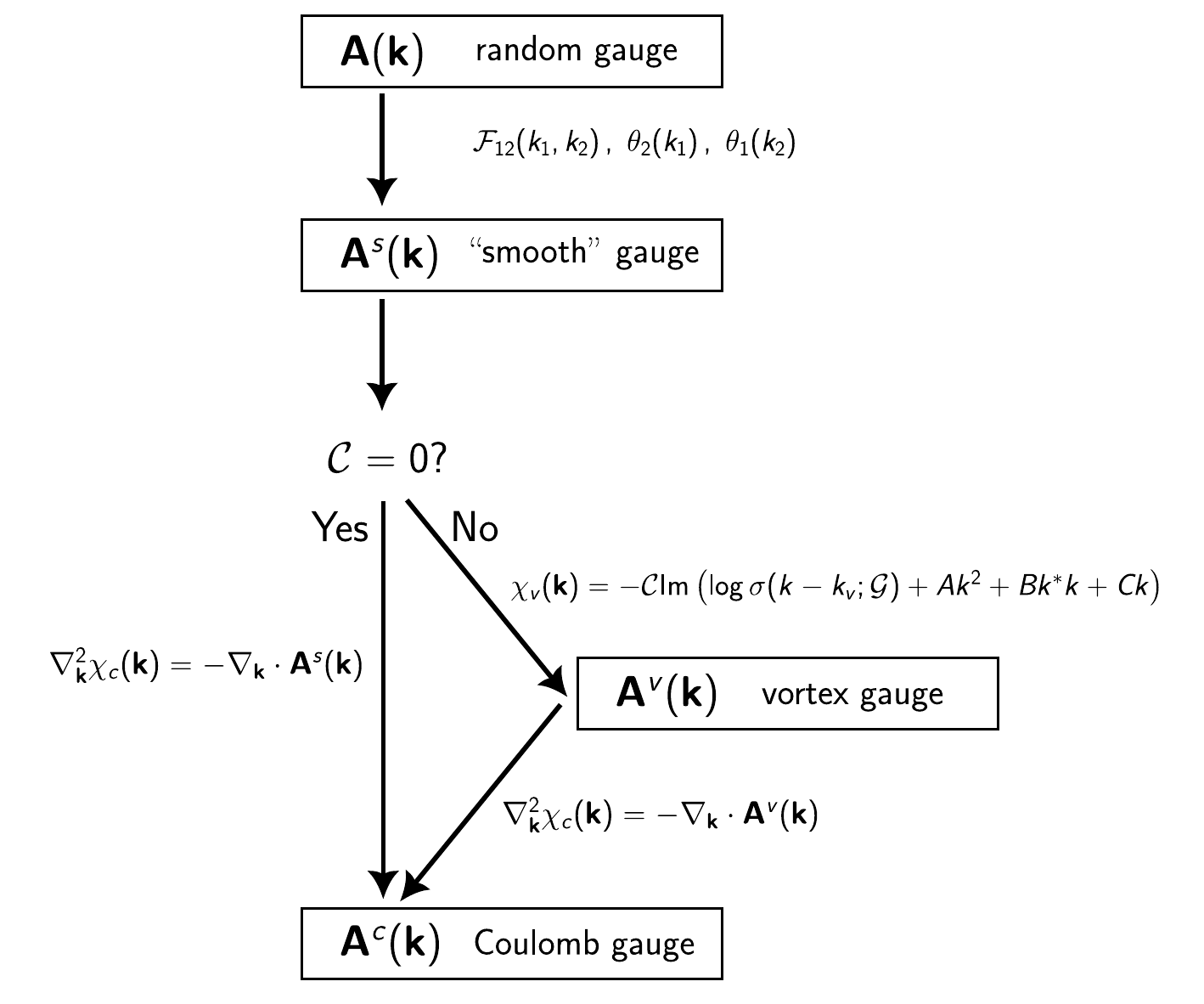}
    \caption{Flowchart showing the gauge fixing procedure for an isolated two dimensional Bloch band. Starting from a random gauge choice, one can compute the gauge-invariant Berry curvature and Wilson loops to obtain a ``smooth'' gauge. If the band has a Chern number, the ``smooth" gauge will have a discontinuity on the ``boundary'' of the Brillouin zone. An additional gauge transformation constructed from a Weierstrass $\sigma$ function can transform this discontinuity into a vortex singularity at any desired position. Finally, the Coulomb gauge choice is obtained by solving a Poisson's equation.
    }
    \label{fig:flowchart}
\end{figure}

In this work, we will address both issues with an algorithm to construct the optimal Wannier functions for generic single band in two dimensions of any non-zero Chern number, which is efficient and does not require any numerical optimization process. 
We directly construct the full gauge transformation from a random gauge to the optimal gauge, by a) applying a smooth gauge transformation introduced in Ref.~\cite{soluyanov_smooth_2012}, followed by b) a vortex gauge transformation and, then, c) solving a Poisson's equation.
Step b) on the vortex gauge transformation is in a similar spirit with -- though is distinct from -- what was recently done in Ref.~\cite{gunawardana2023optimally}.
The flowchart of our procedure is summarized in Fig.~\ref{fig:flowchart}, which will be explained in detail below. 

The optimal Wannier function provides the real-space basis for studying interacting phases in Chern bands.
Specifically, we use the fractional Chern insulator state, or more precisely, the $1/3$ Laughlin state in the kagome lattice model as an example to demonstrate how multi-center exchange interaction terms, generated by the power-law tails of the optimal Wannier functions, can affect the interacting phases in Chern bands. 
We expect these Wannier functions to be useful in understanding other correlated phenomena in generic Chern bands as well.

{\it Gauge fixing.} 
The Wannier function localization functional (WFLF), which quantitatively measures the spread of the Wannier function, is denoted as $F[W]$ and defined as the variance of the position operator \cite{marzari_maximally_2012}. 
For {\it an isolated} energy band gapped from other bands, the WFLF can also be interpreted as a functional of a $U(1)$ gauge transformation defined over the Brillouin zone $F[\chi(\vk)]$. 
In this scenario, finding the functional extrema of the WFLF is equivalent \cite{blount_formalisms_1962} to finding the gauge transformation that results in a Berry connection field satisfying the Coulomb gauge condition:
\begin{equation}\label{eqn:coulomb}
    \nabla_{\vk} \cdot \mathbf{A}(\vk) = 0\,.
\end{equation}
Therefore, the goal of finding the optimal gauge choice of a given Chern band in two dimensions can be summarized as finding the gauge transformation that satisfies Eq.~(\ref{eqn:coulomb}). 
A solution exists for any position of the gauge vortex in the Brillouin zone.

Here we introduce an algorithm that directly constructs such a gauge transformation. The flowchart of our algorithm is summarized in Fig.~\ref{fig:flowchart}. 
We start from solving the Bloch wave functions on a momentum grid over the first Brillouin zone.
The Berry curvature and the non-contractible Wilson loops, which are gauge invariant quantities, can be computed from these Bloch states without any specific gauge fixing.
Using these gauge invariant quantities, one can easily find a smooth gauge choice, denoted as $\mathbf{A}^s(\vk)$ in Fig.~\ref{fig:flowchart}, using a gauge fixing procedure \cite{soluyanov_smooth_2012}.
More specifically, such gauge fixing readily provides a smooth gauge choice over the whole Brillouin zone if the band of interest has a zero Chern number.
As we have shown in Sec.~B 1 in the supplemental material (SM) \cite{supplemental_material}, if the band has a non-zero Chern number, the Bloch state in this gauge is smooth and periodic in $\mathbf{b}_2$ direction, but no longer periodic in $\mathbf{b}_1$: the Bloch state will have a phase jump by $-\mathcal{C}\vk\cdot \mathbf{a}_2$ when $\vk\cdot \mathbf{b}_1$ is increased by $2\pi$.
Here $\mathbf{b}_{1,2}$, $\mathbf{a}_{1,2}$ stand for the reciprocal and Bravais lattice vectors, and $\mathcal{C}$ is the band Chern number.

In order to move this gauge discontinuity into a single vortex, we need a gauge transformation $\chi_v(\vk)$, which is {\it quasiperiodic} such that it ``compensates'' the discontinuity along the Brillouin zone ``boundary'', as presented in Eqs.~(S71-S72) in the SM \cite{supplemental_material}.
Indeed, such condition can be satisfied by utilizing the important properties of the Weierstrass $\sigma$ function \cite{whittaker1920course}, which is {\it quasiperiodic} in the complex plane.
With properly chosen coefficients, the gauge transformation $\chi_v(\vk)$ can be constructed explicitly with the vortex singularity located at any desired position $\vk_v$:
\begin{equation}\label{eqn:vortex-gauge-transformation}
    \chi_v(\vk) = -\mathcal{C}\,{\rm Im}\left(\log \sigma(k-k_v;\mathcal{G}) + Ak^2 + Bk^* k + Ck\right),
\end{equation}
in which $k = k_x + ik_y$ is the complex number representation of momentum point, and $k_v$ is the complex number representation of the vortex position.
Moreover, since $\sigma(z;\mathcal{G})\rightarrow z$ when $z\rightarrow 0$, this gauge transformation can naturally generate a vortex singularity at $\vk_v$.
The constants $A, B$ and $C$, whose expressions are also provided in Sec.~B 2 of the SM \cite{supplemental_material}, are complex numbers which only depend on the shape of the Brillouin zone, and the position of the vortex singularity. 
Applying $\chi_v(\vk)$ to $\mathbf{A}^s(\vk)$, we obtain the ``vortex gauge'' $\mathbf{A}^v(\vk)$ as denoted in Fig.~\ref{fig:flowchart}.

Since the gauge transformation $\chi_v(\vk)$ is constructed explicitly, the vortex gauge Berry connection $\mathbf{A}^v(\vk)$ can be expressed in an analytic form. 
Unfortunately, it does not satisfy Eq.~(\ref{eqn:coulomb}), and thus it is not the optimal gauge choice. 
To meet the Coulomb gauge condition, another gauge transformation $\chi_c(\vk)$ is required. This transformation must satisfy the following Poisson's equation $\nabla_\vk^2\chi_c(\vk) = -\nabla_\vk\cdot \mathbf{A}^v(\vk)$. 
We can prove that the divergence of the vortex gauge Berry connection $\nabla_\vk\cdot \mathbf{A}^v(\vk)$ {\it integrates to zero} over the Brillouin zone, as shown in Sec.~B 3 of the SM \cite{supplemental_material}. 
This guarantees the existence of a smooth solution to $\chi_c(\vk)$ \cite{donaldson2011riemann}, and allows us to construct $\chi_c(\vk)$, by simply solving this Poisson's equation using Fourier transformation:
\begin{equation}\label{eqn:coulomb-gauge-fourier}
    \chi_c(\vk) = \sum_{\mathbf{R}\neq0}\frac{e^{i\vk\cdot \mathbf{R}}}{|\mathbf{R}|^2}\int_{\rm BZ}\frac{d^2k'}{V_{\rm BZ}}\nabla_{\vk'}\cdot\mathbf{A}^v(\vk') e^{-i\vk'\cdot \mathbf{R}} + {\rm const.}
\end{equation}
By applying such gauge transformation to the vortex gauge $\mathbf{A}^v(\vk)$, we obtain a gauge choice $\mathbf{A}^c(\vk)$ that perfectly satisfies Eq.~(\ref{eqn:coulomb}).
As a result, the Wannier function of the Chern band with minimal spread can be constructed by Fourier transforming the Bloch states under this gauge choice.

This construction only requires the knowledge of the Berry curvature and the non-contractible Wilson loops, and no gradient descent algorithm is needed in the process.
We also note that the gauge fixing procedure introduced in this work can be applied to any isolated bands in two dimensions, including both tight binding models and continuum models such as Bistritzer-MacDonald models \cite{Bistritzer2011Moire,Wu2019topological}.

\begin{figure}[t]
    \includegraphics[width=\linewidth]{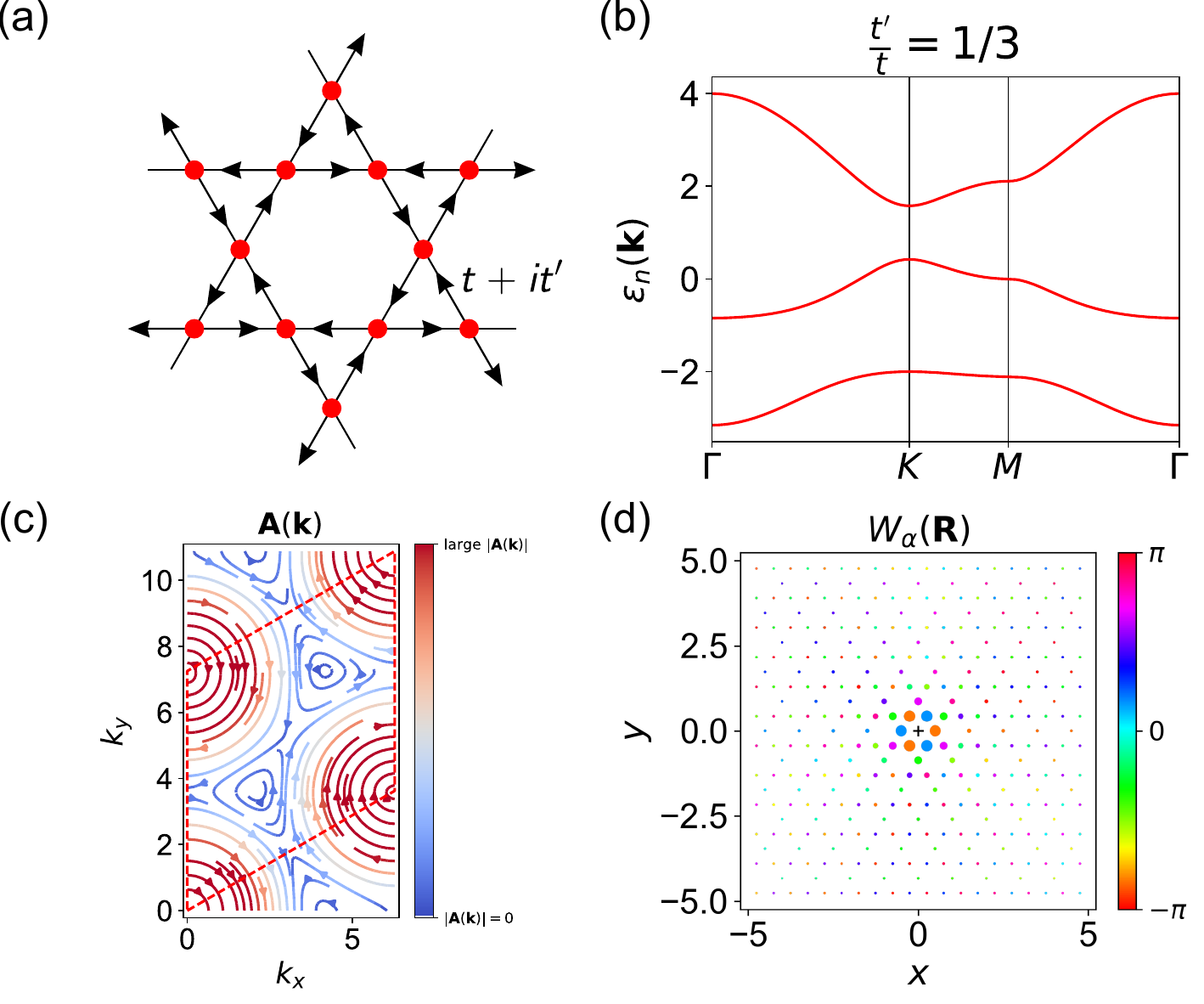}
    \caption{The kagome lattice model. (a) The real space lattice structure. Hoppings along the arrows are given by $t+it'$. (b) The band structure of the kagome lattice model. The Chern number of the lowest band is $\mathcal{C} = 1$. (c) The Berry connection of the lowest Chern band with the vortex located at $\Gamma$ point. (d) The optimal Wannier function of the lowest Chern band with the vortex located at $\Gamma$ point. The size and the color of the circles represents the amplitude and phase of the Wannier function, respectively. The black ``$+$'' symbol stands for the charge center position of this Wannier function.
    The tight binding parameters are chosen as $t'/t=1/3$.
    }
    \label{fig:kagome-summary}
\end{figure}

It is worth noting that the ``ideal Chern bands'' \cite{wang_exact_2021, ledwith2023vortexability}, which are usually considered as the generalization of lowest Landau levels, are naturally written in the optimal gauge, as we will now explain.
The Bloch states of ideal Chern bands can be written as holomorphic/anti-holomorphic functions in the complex momentum space with proper gauge choice, leading to the so-called ``K\"ahler potential'' structure in its Berry connection \cite{claassen_position-momentum_2015, wang_exact_2021, ledwith_fractional_2020}.
Berry connection fields generated from such K\"ahler potential naturally satisfy the Coulomb gauge condition.
Indeed, the magnetic Wannier states of the lowest Landau level introduced in Ref.~\cite{Rashba1997Orthogonal} can be considered as a gauge choice that satisfies Eq.~(\ref{eqn:coulomb}) with a vortex singularity at $\vk_v = (\pi, \pi)$ in the magnetic Brillouin zone.
In other words, the gauge choice constructed via the procedure in Fig.~\ref{fig:flowchart} can also be interpreted as a ``generalization'' of the holomorphic gauge choice in the ideal Chern bands to the case of arbitrary Chern bands.

In Fig.~\ref{fig:kagome-summary}, we show the result of our algorithm applied to the kagome lattice model with complex hoppings, presented in Figs.~\ref{fig:kagome-summary}(a-b), whose lowest band carries a Chern number $\mathcal{C} = 1$ \cite{Tang2011High}.
The Berry connection of this band with the vortex singularity located at the $\Gamma$ point is shown in Fig.~\ref{fig:kagome-summary}(c).
Such choice of the vortex position naturally leads to a more ``symmetric'' Berry connection distribution.
Correspondingly, the Wannier function of this gauge choice is also shown in Fig.~\ref{fig:kagome-summary}(d).
This Wannier state has the charge center located at the center of the hexagonal plaquette.
It resembles the exponentially-localized compact molecular orbitals of a two-orbital kagome metal system \cite{chen2023Metallic} (which, in contrast to the compact localized states in a real-hopping kagome lattice model \cite{Bergman2008Band}, are complete and orthogonal), although the power-law decay of the Wannier function is still present.
The implications of this resemblance are left for a future study.

Further details about the Wannier functions of the kagome lattice model are given in Sec.~C of the SM \cite{supplemental_material}.
There, we also construct and analyze the Wannier functions of another model, one for the twisted bilayer transition metal dichalcogenides (TMD).

{\it Application to a fractionally-filled Chern band in a kagome lattice}.
These optimal Wannier functions can be used as the ``local'' basis for studying the interacting phases in Chern bands. 
Once the density-density interaction terms are projected into these states, multiple types of terms in the Wannier basis can be generated. 
For example, the projected Hamiltonian in a spinless Chern band can be written as:
{\begin{equation}\label{eqn:int}
    \overline{H}_{\rm int} = \frac12 \sum_{\mathbf{R}_0\mathbf{Rdd}'}\mathcal{V}(\mathbf{R};\mathbf{d},\mathbf{d}')w^\dagger_{\mathbf{R}+\mathbf{d}+\mathbf{R}_0}w^\dagger_{\mathbf{d}' + \mathbf{R}_0}w_{\mathbf{R}_0}w_{\mathbf{R} + \mathbf{R}_0},
\end{equation}
in which the sum is over all lattice vectors for $\mathbf{R}, \mathbf{d}, \mathbf{d}'$ and $\mathbf{R}_0$, and $w_\mathbf{R}^\dagger$ indicates the fermion creation operator of the Wannier state centered at unit cell $\mathbf{R}$.
In Fig.~\ref{fig:int}(a), we provide the sketches for the different interaction channels, which contain the direct and exchange channels $\overline{H}_V$ \& $\overline{H}_X$, density assisted hoppings and exchange channels $\overline{H}_{A_h}$ \& $\overline{H}_{A_{ex}}$, as well as the four-center exchange interactions.
Here, we classify these four-center interactions into two types: the ring-exchange terms, $\overline{H}_R$, where the four operators are located at the corners of a plaquette (or equivalently, the four-center terms with the shortest real-space separation), and other generic exchange interactions, $\overline{H}_{\rm extra}$. 
The explicit form of these terms can be found in Sec.~D of the SM \cite{supplemental_material}, especially in Table S1.

\begin{figure}[t]
    \includegraphics[width=\linewidth]{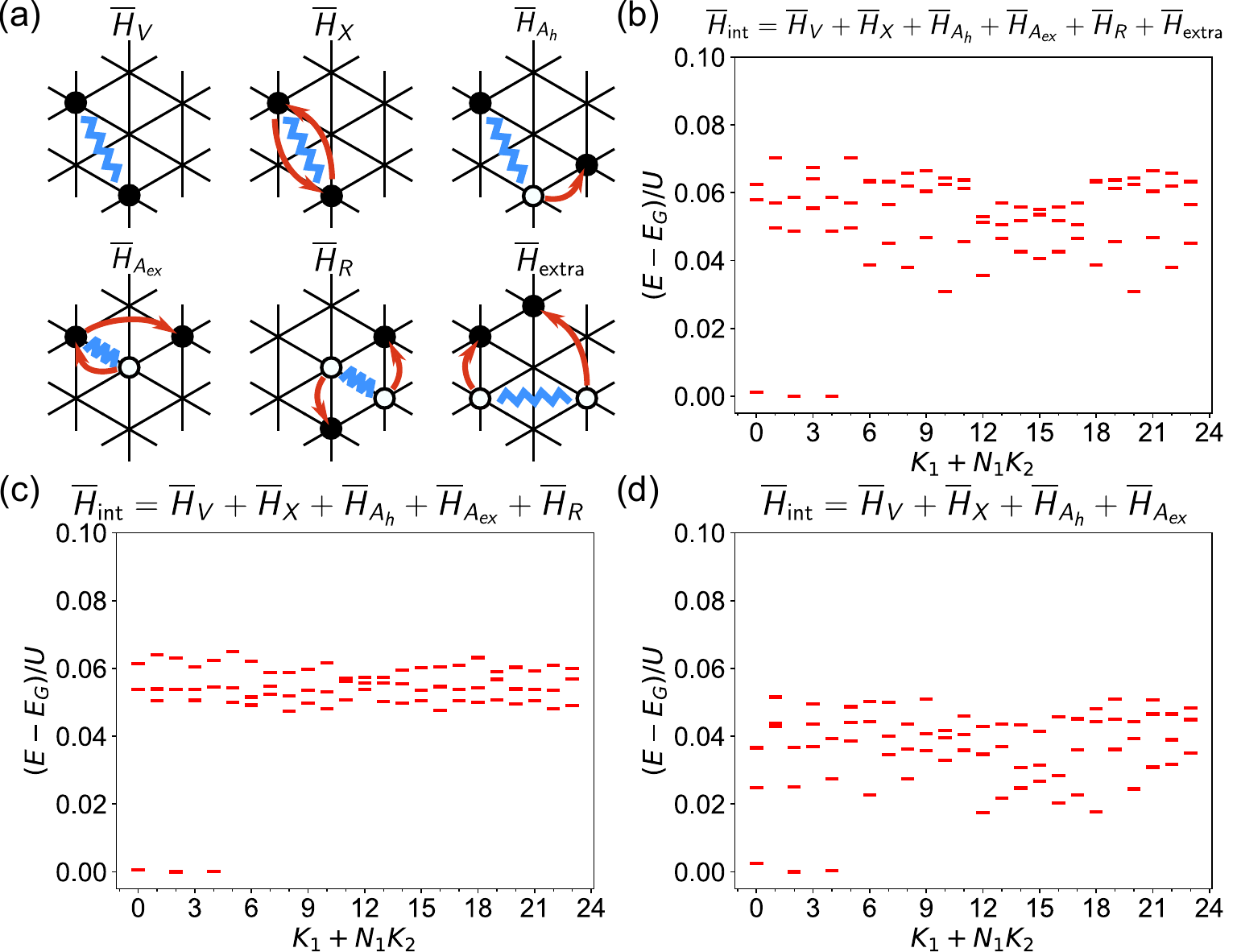}
    \caption{(a) Different interaction channels
    in a spinless Chern band. Filled and empty circles stand for the position of creation and annihilation operators. The blue wavy lines and red arrows represent the vectors $\mathbf{R}$, $\mathbf{d}$ and $\mathbf{d}'$ that are defined in Eq.~(\ref{eqn:int}).
    Their second quantized form are described in the main text and given in Table S1 of the SM
    \cite{supplemental_material}.
    (b) The FCI state is obtained by diagonalizing the interacting Hamiltonian projected into the lowest Chern band of the kagome lattice model. 
    (c) By removing the generic four-center interaction terms, the FCI state can still be obtained.
    (d) After the ring-exchange terms are removed from the Hamiltonian, the energy gap almost vanishes.
    In panels (b-d), the nearest neighbor repulsion strength is taken to be $U/t = 20$.
    }
    \label{fig:int}
\end{figure}

Indeed, the long-range nature of the Wannier states in Chern bands can naturally generate these multi-center exchange terms. 
Our construction of the optimal Wannier functions provides a reasonable basis for studying these multi-center exchange terms in {\it generic} Chern bands, beyond just the lowest Landau levels.
To demonstrate that this optimal Wannier basis can be used as a real space basis, we consider the FCI state in the kagome Chern band, shown in Fig.~\ref{fig:kagome-summary}, as an example. 
Using the optimal Wannier function, we can analyze how different interaction channels affect the stability of the FCI state.
As demonstrated in the energy spectrum in Fig.~\ref{fig:int}(b),
the $1/3$ Laughlin state is obtained as the ground state by exactly diagonalizing the projected interaction Hamiltonian when all the terms in the projected Hamiltonian are included 
(cf., Ref.~\cite{wu2012zoology}).
If the generic four-center exchange terms $\overline{H}_{\rm extra}$ are removed from the Hamiltonian, the FCI state persists as the ground state, as shown in Fig.~\ref{fig:int}(c). 
However, if one further truncates the ring-exchange terms $\overline{H}_R$ from the Hamiltonian, the FCI state can no longer be obtained, which is indicated by the strongly reduced energy gap shown in Fig.~\ref{fig:int}(d). 
The calculated particle entanglement spectrum, given in Sec.~D 2 of the SM \cite{supplemental_material}, also indicates the destruction of the Laughlin states in the absence of the ring-exchange terms.

The results in Figs.~\ref{fig:int}(b-d) demonstrate that the ring-exchanges are responsible for stabilizing a {\it fermionic} Laughlin state from Chern bands (beyond Landau levels).
Thus, our work sheds new light on the mechanism for stablizing the FCI states.

{\it Discussion.} 
Several remarks are in order. Firstly, we have focused on the case of a single isolated Chern band. We leave the extension of our method to multi-band systems for the future.

Secondly, to put our results on the mechanism for the FCIs in perspective, traditionally, in the Hubbard-Kanamori Hamiltonian \cite{Kanamori1963, Georges2013Strong}, only the direct and exchange channels (i.e., $\overline{H}_V$ and $\overline{H}_X$) are considered.
Many non-trivial phenomena in interacting Chern bands, including the FCI states, cannot be realized by the Hubbard-Kanamori type Hamiltonian.
On the other hand, parent Hamiltonians with Kalmeyer-Laughlin type chiral spin liquids as the exact ground states, which are also equivalent to bosonic fractional quantum Hall states, have been proposed \cite{Kalmeyer1987,Schroeter2007Spin,Nielsen2012Laughlin}.
These Hamiltonians contain long-range chiral coupling terms with the form of, {\it e.g.}, $\mathbf{S}_i\cdot(\mathbf{S}_j\times \mathbf{S}_k)$ in the spin representation, or equivalently, multi-center exchange interactions in the boson representation. 
Additionally, the importance of the assisted hoppings and ring-exchange terms have been emphasized in the context of {\it bosons} in the lowest Landau level \cite{Burkov2010fractional, Panfilov2016chiral}.
Our results show that the multi-center exchange terms are important in the present context of interacting electron systems as well. 

Thirdly, the proposed procedure for constructing Wannier states of Chern bands opens the door to the study of other interaction effects.
For example, as we have addressed in this work, the Coulomb gauge is deeply related to the holomorphic gauge in ideal Chern bands. 
Exploring how the optimal Wannier states and their corresponding interaction elements deform from ideal Chern bands to generic Chern bands will be instructive.
Another perspective is applying the optimal Wannier states in spinful systems, which have been considered through time-reversal symmetry breaking Wannier states that are exponentially localized \cite{Setty2021Electron}.
Spinful topological bands with stronger interactions have also been explored in a symmetry-preserving way, in terms of compact molecular orbitals \cite{chen2023Metallic,Chen_emergent_2024,HuSciAdv2023}.

In relation to the last point, it is intriguing to see that the core region of the optimally localized Wannier orbitals constructed here resembles those of the exponentially-localized compact molecular orbitals (comparing Fig.~\ref{fig:kagome-summary}(d) to Figs.~2(d) and S4(b)  of Ref.~\cite{chen2023Metallic}). 
This suggests the possibility of smoothly connecting the physics in the correlation regime of isolated Chern bands and the regime in which these flat bands are coupled to wider bands.

Finally, the recently studied moir\'e materials such as the twisted multilayer TMDs are expected to host narrow Chern bands near the Fermi level \cite{Wu2019topological}. 
Various types of non-trivial correlated phenomena have been studied in these systems, including fractional Chern insulators 
\cite{Cai2023Signatures,Zeng2023Thermodynamic,Park2023Observation,Kang2024Evidence} (see also Refs.~\onlinecite{Xie2021Fractional,Lu2024Fractional}).

We expect the construction of optimal Wannier functions will be important to the understanding of these realistic systems.
With that in mind, we have briefly discussed the optimal Wannier functions for the twisted bilayer TMD in Sec.~C 2 of the SM \cite{supplemental_material}.

{\it Summary.}
We have advanced a general procedure to construct the optimally localized Wannier functions for an isolated Chern band. 
We have used these Wannier functions to study interaction effects in Chern bands, demonstrating that multi-center exchange terms are crucial for stabilizing fractional Chern insulators in 
partially filled Chern bands.
Our findings set the stage to study the competition between fractional Chern insulators and other correlated phases.
Moreover, our results raise the prospect for smoothly connecting the correlation physics of isolated Chern flat bands and that arising in the coupled flat and wide bands, thereby expanding the realization and understanding of new physics from the interplay between strong correlations and topology in flat band systems.

{\it Note added}: After the completion of this manuscript, a recent work discussing a complementary construction of the Wannier function from Chern bands based on an alternative method, using Bloch states with properly chosen normalization functions, became available \cite{Li2024Constraints}.

{\it Acknowledgments.} We thank Gabriel Aeppli, Pengcheng Dai, Yichen Fu, Patrick Ledwith, Silke Paschen, Chandan Setty, Shouvik Sur, 
Roser Valent\'\i, Jie Wang, Yonglong Xie and Ming Yi for useful discussions.
Work at Rice has primarily been supported by the U.S. DOE, BES, under Award No. DE-SC0018197 (Wannier construction, F.X., Y.F., L.C.), by the Air Force Office of Scientific Research under Grant No. FA9550-21-1-0356 (interaction effect, F.X., Y.F., L.C.), by the Robert A. Welch Foundation Grant No. C-1411 (Q.S.) and the Vannevar Bush Faculty Fellowship ONR-VB N00014-23-1-2870 (Q.S.).
J.C. acknowledges the support of the National Science Foundation under Grant No. DMR-1942447, support from the Alfred P. Sloan Foundation through a Sloan Research Fellowship and the support of the Flatiron Institute, a division of the Simons Foundation.
The majority of the computational calculations have been performed on the Shared University Grid at Rice funded by NSF under Grant EIA-0216467, a partnership between Rice University, Sun Microsystems, and Sigma Solutions, Inc., the Big-Data Private-Cloud Research Cyberinfrastructure MRI-award funded by NSF under Grant No. CNS-1338099, and the Extreme Science and Engineering Discovery Environment (XSEDE) by NSF under Grant No. DMR170109. 
Q.S. acknowledges the hospitality of the Aspen Center for Physics, which is supported by the National Science Foundation under Grant No. PHY-2210452.

\bibliography{reference.bib}
\bibliographystyle{apsrev4-2}

\clearpage

\onecolumngrid
\beginsupplement
\section*{Supplemental Materials}

\tableofcontents

\section{Basics}\label{sec:basics}

In this section, we establish notations for the Bloch states, the Berry connection and Berry curvature on discrete momentum mesh.

\subsection{Notations for real and reciprocal lattices}

The coordinates of a two dimensional periodic lattice site can be represented using the Bravais lattice basis vectors $\mathbf{a}_1, \mathbf{a}_2$:
\begin{equation}
    \mathbf{R} = R_1 \mathbf{a}_1 + R_2 \mathbf{a}_2\,,
\end{equation}
in which $R_1, R_2 \in \mathbb{Z}$. Accordingly, the basis vectors of the reciprocal (momentum) space can also be represented as follows:
\begin{equation}
    \mathbf{G} = G_1 \mathbf{b}_1 + G_2 \mathbf{b}_2\,,
\end{equation}
where $G_1, G_2 \in \mathbb{Z}$. The basis vectors in real space and reciprocal space satisfy the following relationship:
\begin{equation}
    \mathbf{a}_i \cdot \mathbf{b}_j = 2 \pi \delta_{ij}\,.
\end{equation}
The unit cell in the reciprocal space spanned by $\mathbf{b}_1, \mathbf{b}_2$ is the Brillouin zone. Any momentum point $\vk$ in the first Brillouin zone can be represent by two real numbers $k_1, k_2 \in [0, 2\pi)$:
\begin{equation}
    \vk = \frac{k_1}{2\pi}\mathbf{b}_1 + \frac{k_2}{2\pi}\mathbf{b}_2.
\end{equation}
If the lattice is finite and periodic along both the directions with $N_1$ and $N_2$ unit cells, each momentum point can also be represented by two integers $K_1 = 0, 1,\cdots N_1 - 1$ and $ K_2 = 0, 1, \cdots N_2 - 1$:
\begin{equation}\label{eqn:discrete-momentum}
    \vk = \frac{K_1}{N_1} \mathbf{b}_1 + \frac{K_2}{N_2} \mathbf{b}_2\,.
\end{equation}
The real number representation of these discrete momenta points are given by $k_1 = 2 \pi K_1/N_1$ and $k_2 = 2 \pi K_2/N_2$. An example of discretized momentum space can be found in Fig.~\ref{fig:kmesh}(a). 

Note that we will always use $k_1, k_2$ and $G_1, G_2$ to represent the components of $\vk$ and $\mathbf{G}$ under the reciprocal lattice basis. On the other hand, symbols including $k_x, k_y$ and $G_x, G_y$ represent the momentum components under the Cartesian basis:
\begin{align*}
    k_x &= \vk \cdot \hat{\mathbf{x}}\,, k_y = \vk \cdot \hat{\mathbf{y}}\,;\\
    G_x &= \mathbf{G} \cdot \hat{\mathbf{x}}\,, G_y = \mathbf{G} \cdot \hat{\mathbf{y}} \,.
\end{align*}

\begin{figure}
    \centering
    \includegraphics[width=0.75\linewidth]{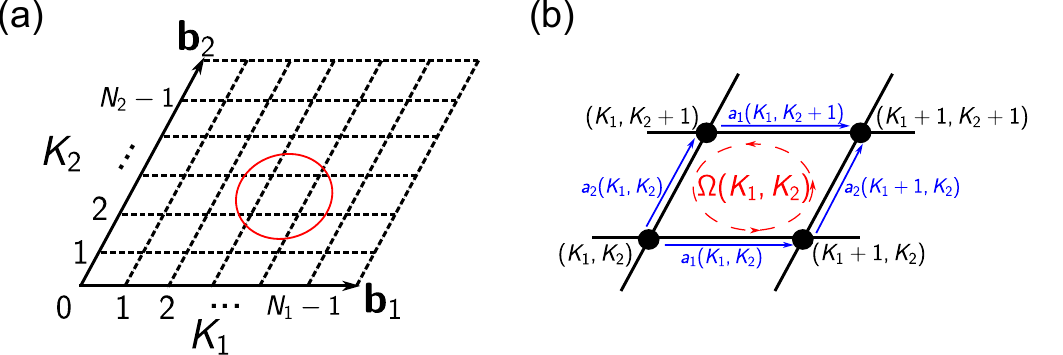}
    \caption{(a) The Brillouin zone is discretized into $N_1 \times N_2$ points. (b) The Berry connection $a_i(K_1, K_2)$ is defined along the links between neighboring momentum points. The Berry curvature $\Omega(K_1, K_2)$ is defined in each plaquette in momentum space. The orientation of $a_i$ and $\Omega$ are indicated by the colored arrows in the figure.}
    \label{fig:kmesh}
\end{figure}

\subsection{Bloch Hamiltonian}
\subsubsection{Tight-binding models}
Tight-binding models can be defined on such momentum mesh. Generally, the Hamiltonian of a tight-binding model has the following form:
\begin{equation}
    H_0 = \sum_{\mathbf{R}, \mathbf{R}', \alpha\beta}t_{\alpha\leftarrow\beta}(\mathbf{R})c^\dagger_{\mathbf{R} + \mathbf{R}',\alpha}c_{\mathbf{R}',\beta}\,,
\end{equation}
in which $c_{\mathbf{R}',\beta}$ stands for the fermion/boson annihilation operator corresponding to the one-body states from sublattice $\beta$ in the unit cell centered at $\mathbf{R}'$.
The quantity $t_{\alpha\leftarrow\beta}(\mathbf{R}) \in \mathbb{C}$ is the hopping strength from the orbital $\beta$ in an arbitrary unit cell located at $\mathbf{R}_0$, to the orbital $\alpha$ in the unit cell located at $\mathbf{R} + \mathbf{R}_0$.
The fermionic/bosonic operators defined in momentum space are related to these real space operators via discrete Fourier transformation:
\begin{equation}
    c^\dagger_{\vk,\alpha} = \frac{1}{\sqrt{N_1N_2}}\sum_\mathbf{R}c^\dagger_{\mathbf{R},\alpha}e^{i\vk\cdot (\mathbf{R} + \bm{\tau}_\alpha)}\,,
\end{equation}
in which $\bm{\tau}_\alpha$ stands for the (relative) position of the sublattice $\alpha$.
The tight-binding Hamiltonian defined with operators $c^\dagger_{\mathbf{R},\alpha}$ can be Fourier transformed into the following form:
\begin{align}
    H_0 &= \sum_{\mathbf{k}\in{\rm BZ}}\sum_{\alpha\beta} h_{\alpha\beta}(\vk)c^\dagger_{\vk,\alpha}c_{\vk,\beta}\,,\\
    h_{\alpha\beta}(\vk) &= \sum_{\mathbf{R}}t_{\alpha\leftarrow\beta}(\mathbf{R}) e^{-i\vk\cdot(\mathbf{R} + \bm{\tau}_\alpha - \bm{\tau}_\beta)}\,,\label{eqn:hopping-fourier}
\end{align}
in which the momentum point set $\rm BZ$ is determined by aforementioned condition Eq.~(\ref{eqn:discrete-momentum}). 
Note that if there are orbitals at different positions in each unit cell, the Hamiltonian $h(\vk)$ will not be periodic in the reciprocal space. Instead, the embedding matrix $V_\mathbf{G}$ is required:
\begin{align}
    h(\vk + \mathbf{G}) &= V_\mathbf{G} h(\vk) V^{-1}_{\mathbf{G}}\,,\\
    (V_\mathbf{G})_{\alpha\beta} &= \delta_{\alpha\beta} e^{-i\mathbf{G}\cdot \bm{\tau}_\alpha}\,.\label{eqn:emb-tb}
\end{align}

One-body eigenstates can be computed by diagonalizing the Bloch Hamiltonian $h(\vk)$. We will denote the $n$-th eigenvector of $h(\vk)$ as $u_{\alpha, n}(\vk)$, and the corresponding eigenvalue as $\varepsilon_{n}(\vk)$:
\begin{equation}\label{eqn:bloch-eigen}
    \sum_{\beta}h_{\alpha\beta}(\vk) u_{\beta, n}(\vk) = \varepsilon_n(\vk) u_{\alpha, n}(\vk)\,.
\end{equation}
The eigenvalues $\varepsilon_n(\vk)$ provide the dispersion of the energy bands, and the creation operator of the Bloch state in the $n$-th energy band is given by:
\begin{align}
    \gamma^\dagger_{\vk, n} &= \sum_\alpha u_{\alpha, n}(\vk) c^\dagger_{\vk,\alpha}\nonumber\\
    &=\frac{1}{\sqrt{N_1N_2}}\sum_{\mathbf{R}\alpha}u_{\alpha,n}(\vk) e^{i\vk\cdot(\mathbf{R} + \bm{\tau}_\alpha)} c^\dagger_{\mathbf{R},\alpha}\,,\label{eqn:band-operator}
\end{align}
and $u_{\alpha,n}(\vk)$ can be regarded as the periodic part of the Bloch wave function.

We note that the Bloch Hamiltonian $h(\vk)$ is generically not periodic under the momentum translation $\vk \rightarrow \vk + \mathbf{G}$, and this could lead to non-periodic eigenvectors $u_{\alpha, n}(\vk) \neq u_{\alpha, n}(\vk + \mathbf{G})$. However, $u_{\alpha, n}(\vk)$ and $u_{\alpha, n}(\vk + \mathbf{G})$ should still represent the same quantum state in the one-body Hilbert space. 
One can easily prove that the Bloch band creation operators are periodic in the reciprocal space $\gamma^\dagger_{\vk + \mathbf{G},n} = \gamma^\dagger_{\vk,n}$, once the following equation is enforced:
\begin{equation}\label{eqn:tb-emb-translation}
    u_{n}(\vk + \mathbf{G}) = V_{\mathbf{G}} u_{n}(\vk)\,.
\end{equation}

\subsubsection{Continuum models}
Another type of widely studied Bloch systems are continuum models subject to a periodic potential. Bistritzer-MacDonald model for twisted bilayer graphene falls into this category \cite{Bistritzer2011Moire}. The formulation of Bloch states in continuum models is different than their tight-binding counter parts, yet they still share some similarities.

Hamiltonians of continuum models are usually represented in the plane wave basis:
\begin{align}
    H_0 &= \sum_{\vk\in{\rm BZ}}\sum_{\mathbf{G}\mathbf{G}'\in\mathcal{G}} h_{\mathbf{G},\mathbf{G}'}(\vk) c^\dagger_{\vk - \mathbf{G}} c_{\vk - \mathbf{G}'}\,,\\
    h_{\mathbf{G},\mathbf{G}'}(\vk) &= \varepsilon(\vk - \mathbf{G})\delta_{\mathbf{G},\mathbf{G}'} + U_{\mathbf{G} - \mathbf{G}'}\,.\label{eqn:def-pw-hamiltonian}
\end{align}
Here $\varepsilon(\vk)$ is the dispersion relation without periodic potential, $\mathbf{G},\mathbf{G}'$ are the reciprocal lattice vectors, and $U_\mathbf{G}$ is the Fourier transformation of the periodic potential.
Fermionic/bosonic operator $c^\dagger_{\vk-\mathbf{G}}$ creates a plane wave state with momentum $\vk - \mathbf{G}$:
\begin{equation}
    c^\dagger_{\vk - \mathbf{G}} = \frac{1}{\sqrt{N_1N_2\Omega_c}}\int d^2r\, e^{i(\vk - \mathbf{G})\cdot \mathbf{r}}\psi^\dagger(\mathbf{r})\,,
\end{equation}
in which $\psi^\dagger(\mathbf{r})$ is the fermion/boson field operator at position $\mathbf{r}$, and $\Omega_c$ is the unit cell volume.

Diagonalizing $h(\vk)$ yields the band structure and the corresponding eigenvectors $u_{\mathbf{Q},n}(\vk)$.
Accordingly, Bloch states can be created by the following operator:
\begin{align}
    \gamma^\dagger_{\vk,n} =& \sum_{\mathbf{G}} u_{\mathbf{G},n}(\vk) c^\dagger_{\vk - \mathbf{G}}\nonumber\\
    =& \frac{1}{\sqrt{N_1N_2 \Omega_c}}\int d^2r\,e^{i\vk\cdot\mathbf{r}} \sum_{\mathbf{G}}u_{\mathbf{G},n}(\vk)e^{-i\mathbf{G}\cdot \mathbf{r}} \psi^\dagger(\mathbf{r})\,.
\end{align}
$u_{\mathbf{Q},n}(\vk)$ is also the Fourier transformation of the periodic part of the Bloch wave function. 

The Bloch Hamiltonian $h(\vk)$ in plane wave basis is not a periodic function of $\vk$. Instead, one can show that its embedding matrix is given by:
\begin{equation}\label{eqn:pw-emb}
    \left(V_{\mathbf{G}''}\right)_{\mathbf{G},\mathbf{G}'} = \delta_{\mathbf{G} - \mathbf{G}', \mathbf{G}''}\,.
\end{equation}
Although $V_{\mathbf{G}''}$ is no longer diagonal in this case, we can still guarantee that a Bloch band creation operator of a continuum model is periodic in reciprocal space by enforcing Eq.~(\ref{eqn:tb-emb-translation}).

\subsection{Wannier function}

In this manuscript, we focus on the single band case, meaning that the band of interest is gapped from other bands, and each state in this band can be labeled solely by a momentum quantum number $\vk$.
The Bloch states can be Fourier transformed into the \emph{Wannier states} $|W_n(\mathbf{R})\rangle = w^\dagger_{\mathbf{R},n}|0\rangle$:
\begin{align}
    w^\dagger_{\mathbf{R},n}  =& \frac{1}{\sqrt{N_1N_2}}\sum_{\vk}\gamma_{\vk,n}^\dagger e^{-i\vk\cdot\mathbf{R}}\,.
\end{align}
In both tight-binding and continuum models, Wannier states can be computed through a Fourier transformation of their respective eigenvectors $u_n(\vk)$, although employing a slightly different approach. For tight binding models, the Wannier function is defined on discrete sublattice points:
\begin{align}
    w^\dagger_{\mathbf{R},n} =& \sum_{\mathbf{R}'\alpha} W_\alpha(\mathbf{R}' - \mathbf{R}) c^\dagger_{\mathbf{R}',\alpha} \,,\label{eqn:def-tb-wannier}\\
    W_\alpha(\mathbf{R}' - \mathbf{R}) =& \frac{1}{N_1N_2}\sum_{\vk}u_{\alpha,n}(\vk)e^{i\vk\cdot(\mathbf{R}' - \mathbf{R} +\bm{\tau}_\alpha)} \,.\label{eqn:wannier-tb}
\end{align}
And for continuum models, the Wannier function is defined in continuous real space:
\begin{align}
    w^\dagger_{\mathbf{R},n} &= \int d^2r\, W(\mathbf{r} - \mathbf{R}) \psi^\dagger(\mathbf{r}) \,,\\
    W(\mathbf{r}) &= \frac{1}{N_1N_2\sqrt{\Omega_c}}\sum_{\vk,\mathbf{G}}u_{\mathbf{G},n}(\vk)e^{i(\vk-\mathbf{G})\cdot \mathbf{r}}\,,\label{eqn:wannier-pw}
\end{align}

With Wannier function definition provided, one can also use the \emph{Wannier function localization functional} (WFLF) to quantitatively describe its spread, which is defined in the following form \cite{marzari_maximally_2012, blount_formalisms_1962}:
\begin{equation}\label{eqn:def-wflf}
    F[W] = \langle W_n(\mathbf{0})| \mathbf{\hat{r}}^2 |W_n(\mathbf{0})\rangle - |\langle W_n(\mathbf{0}) | \mathbf{\hat{r}} | W_n(\mathbf{0}) \rangle |^2\,.
\end{equation}
The one-body position operators $\mathbf{\hat{r}}$ in tight-binding models and continuum models are given by Eqs.~(\ref{eqn:def-r-tb}) and (\ref{eqn:def-r-pw}), respectively.
\begin{align}
    \mathbf{\hat{r}} &= \sum_{\mathbf{R}', \alpha} (\mathbf{R'} + \bm{\tau}_\alpha) c^\dagger_{\mathbf{R}', \alpha} |0\rangle \langle 0 | c_{\mathbf{R}', \alpha}\,,\label{eqn:def-r-tb}\\
    \mathbf{\hat{r}} &= \int d^2r'\, \mathbf{r}'\psi^\dagger(\mathbf{r}')|0\rangle \langle 0|\psi(\mathbf{r}')\,.\label{eqn:def-r-pw}
\end{align}

The phase of the eigenvectors $u_{alpha,n}(\vk)$ is not uniquely defined at each $\vk$ and, in particular, numerical diagonalization returns an arbitrary phase. The following {\it gauge transformation} leaves the physical state unchanged:
\begin{equation}
    u_{\alpha, n}(\vk) \rightarrow e^{i\chi(\vk)}u_{\alpha, n}(\vk)\,.
\end{equation}
Observable quantities, such as the charge center of a trivial band, or the Hall conductance of a topological band do not depend on the gauge choice. However, the Wannier function of the corresponding band strongly depends on the gauge choice the of Bloch states. 
The gauge choice of the Bloch functions can affect the value of $F[W]$, and the \emph{maximally localized Wannier functions} \cite{marzari_maximally_2012} for trivial bands corresponds to a special gauge choice.
Hence, choosing a proper gauge for Bloch states is still meaningful, especially when investigating the real space characteristics of a Bloch band.

\subsection{Connection, curvature and Wilson loop}
Berry connection and Berry curvature are the quantities which characterize the change of the Bloch states over the Brillouin zone. We first define these quantities using discrete momentum mesh, and the generalization to continuum limit $N_1, N_2 \rightarrow \infty$ is straightforward. 

We assume that the momentum mesh density is sufficiently high, ensuring that the Bloch state eigenvectors from neighboring momentum points exhibit significant overlaps:
\begin{equation}\label{eqn:dense-condition}
    |u^\dagger_n(\vk) u_n(\vk+\Delta \vk)| \sim 1\,,
\end{equation}
in which $u_n(\vk) = [u_{1,n}(\vk), u_{2,n}(\vk), \cdots]^{\rm T}$ stands for the $n$-th eigenvector of Bloch Hamiltonian $h(\vk)$, and the indices $\alpha=1,2,\cdots$ stand for either sublattices in tight binding models, or plane wave basis in continuum models.
Therefore, the information of the gauge choice is mostly encoded in the phase of the inner products between Bloch states. As an example, the Berry connection vectors $a_i$ are defined along the links between neighboring points, which are labeled by the blue arrows in Fig.~\ref{fig:kmesh}(b): 
\begin{align}
    e^{ia_1(K_1, K_2)} &= \frac{u^\dagger_n(K_1, K_2) u_n(K_1 + 1, K_2)}{|u^\dagger_n(K_1, K_2) u_n(K_1 + 1, K_2)|}\,,\\
    e^{ia_2(K_1, K_2)} &= \frac{u^\dagger_n(K_1, K_2) u_n(K_1, K_2 + 1)}{|u^\dagger_n(K_1, K_2) u_n(K_1, K_2 + 1)|}\,.
\end{align}
The Berry connection vectors are not gauge invariant. Under the gauge transformation $e^{i\chi(\vk)}$, these quantities will transform as follows:
\begin{align}
     a_1(K_1, K_2) &\rightarrow a_1(K_1, K_2) + \chi(K_1 + 1, K_2) - \chi(K_1, K_2)\,,\label{eqn:connection-gauge-1}\\
     a_2(K_1, K_2) &\rightarrow a_2(K_1, K_2) + \chi(K_1, K_2 + 1) - \chi(K_1, K_2)\,.\label{eqn:connection-gauge-2}
\end{align}

If $a_1$ and $a_2$ scale as $a_1 \sim {1}/{N_1}$ and $a_2\sim{1}/{N_2}$ when approaching to the continuum limit $N_1, N_2 \rightarrow \infty$ under certain gauge choice, we are able to associate the discrete Berry connection components to a Berry connection (dual) vector $\mathbf{A}(\vk)$ as follows:
\begin{align}
    \mathbf{A}(\vk) &= -i u^\dagger_n(\vk)\nabla_\vk u_n(\vk)\,,\label{eqn:continuum-connection-1}\\
    a_1(K_1, K_2) &\leftrightarrow \frac{1}{N_1}\mathbf{A}(\vk)\cdot \mathbf{b}_1 \leftrightarrow A_1(k_1, k_2) dk_1  \,, \label{eqn:continuum-connection-2}\\
    a_2(K_1, K_2) &\leftrightarrow \frac{1}{N_2}\mathbf{A}(\vk)\cdot \mathbf{b}_2 \leftrightarrow A_2(k_1, k_2) dk_2 \,.\label{eqn:continuum-connection-3}
\end{align}
Under a smooth gauge transformations $e^{i\chi(\vk)}$, these components will transform as:
\begin{equation}
    A_i(k_1, k_2) \rightarrow A_i(k_1, k_2) + \frac{\partial \chi}{\partial k_i}\,.
\end{equation}
Alternatively, if $a_1$ or $a_2$ approaches to a non-zero constant value when $N_1, N_2\rightarrow \infty$, there will be a sudden phase jump in the gauge choice of the Bloch states, and the corresponding Berry connection $\mathbf{A}(\vk)$ in the continuum limit will diverge. 

The Berry curvature is defined in each plaquette of the momentum mesh, which is shown by the red circle in the middle of Fig.~\ref{fig:kmesh}(b). The expression of the curvature $\Omega(K_1, K_2)$ of the $n$-th band is given by the following equation:
\begin{align}\label{eqn:def-curvature}
    \Omega(K_1, K_2)=& {\rm arg}\Big{(} u^\dagger_n(K_1, K_2)u_n(K_1 + 1, K_2) u^\dagger_n(K_1 + 1, K_2) u_n(K_1 + 1, K_2 + 1) \nonumber \\
    & \times u^\dagger_n(K_1 + 1, K_2 + 1) u_n(K_1, K_2 + 1)  u^\dagger_n(K_1, K_2 + 1) u_n(K_1, K_2)\Big{)}\,.
\end{align}
Practically, the imaginary part of the Bloch states products in Eq.~(\ref{eqn:def-curvature}) is $\ll 1$ when the band of interest is fully gapped, and when the momentum mesh is dense enough such that Eq.~(\ref{eqn:dense-condition}) is satisfied. Thus, we choose the branch of $\rm arg$ function closest to $0$, and this choice will lead to a smooth curvature over the whole Brillouin zone. With the increasing of $N_1$ and $N_2$, the value of $\Omega$ will scale as $\sim 1/(N_1 N_2)$. Similar to $\mathbf{A}(\vk)$, we can also define the Berry curvature as a rank-2 tensor in the limit $N_1, N_2\rightarrow \infty$:
\begin{align}
    \mathcal{F}_{12}(k_1, k_2) &= \frac{\partial A_2(k_1, k_2)}{\partial k_1} - \frac{\partial A_1(k_1, k_2)}{\partial k_2}\,,\label{eqn:continuum-curvature}\\
    \Omega(K_1, K_2) &\leftrightarrow \mathcal{F}_{12}(k_1, k_2)dk_1 dk_2\,.
\end{align}

Unlike the Berry connection, the curvature is clearly gauge invariant, since each Bloch state and its conjugate appeared once in the definition of $\Omega(K_1, K_2)$. Obviously, any gauge transformation will not show up. Summing the Berry curvature over the whole Brillouin zone gives us the Chern number $\mathcal{C} \in \mathbb{Z}$:
\begin{equation}
    \mathcal{C} = \frac{1}{2\pi}\sum_{K_1=0}^{N_1 - 1}\sum_{K_2 = 0}^{N_2 - 1}\Omega(K_1, K_2)\,.
\end{equation}
The Chern number $\mathcal{C}$ can always be defined for a 2D energy band, which is gapped from other bands. When $\mathcal{C} \neq 0$, the band is usually called a {\it Chern band}.

The curvature $\Omega(K_1, K_2)$ can be interpreted as small Wilson loops, or the ``loop integral'' of $a_i$ field, around each plaquette in the momentum mesh. Similarly, we can also define the Wilson loops as the ``integral'' of $a_i$ along the two non-contractible paths of the Brillouin zone torus: 
\begin{align}
    W_1(K_2) = e^{i\theta_1(K_2)} &= \prod_{K_1=0}^{N_1 - 1} e^{ia_1(K_1, K_2)}\,,\label{eqn:def-wilson-1}\\
    W_2(K_1) = e^{i\theta_2(K_1)} &= \prod_{K_2=0}^{N_2 - 1} e^{ia_2(K_1, K_2)}\,.\label{eqn:def-wilson-2}
\end{align}
The Wilson loops are also gauge invariant for a similar reason as that which applies to the Berry curvature. The exponent of the Wilson loops $\theta_1(K_2)$ and $\theta_2(K_1)$ are only well defined $\rm mod~2\pi$. However, we are able to ensure that $\theta_1$ is a continuous function when $K_2$ continuously increased from $0$ to $N_2$ if the branch choice of $\theta_1$ is chosen properly. Since $e^{i\theta_1(0)} = e^{i\theta_1(N_2)}$ due to the periodicity of the Brillouin zone, the value of $\theta_1$ changes by an integer multiple of $2\pi$. This integer is indeed the Chern number:
\begin{equation}\label{eqn:wilson-period-1}
    \theta_1(N_2) - \theta_1(0) = -2\pi \mathcal{C}\,.
\end{equation}
A similar conclusion applies to the Wilson loop along the $\mathbf{b}_2$ axis:
\begin{equation}\label{eqn:wilson-period-2}
    \theta_2(N_1) - \theta_2(0) = 2\pi \mathcal{C}\,.
\end{equation}
Therefore, we shall define the values of $\theta_1(K_2)$ and $\theta_2(K_1)$ by ensuring their continuity.
The Wilson loops can also be written as continuous functions of $k_1, k_2$ in the limit $N_1, N_2 \rightarrow \infty$:
\begin{align}
    W_1(k_2) &= e^{i\theta_1(k_2)} = \mathcal{P}\exp\left(i \int_0^{2\pi} dk_1\,A_1(k_1, k_2)\right)\,,\\
    W_2(k_1) &= e^{i\theta_2(k_1)} = \mathcal{P}\exp\left(i\int_0^{2\pi} dk_2\,A_2(k_1, k_2)\right)\,,
\end{align}
in which $\mathcal{P}$ stands for path-ordering. Wilson loops $W_1(k_2), W_2(k_1)$ and Berry curvature $\mathcal{F}_{12}$ depict the variations, and more importantly, the topological ``winding'' properties of the Bloch wave functions in the momentum space.

We also note that the Berry connection $A_1, A_2$ in Eqs.~(\ref{eqn:continuum-connection-2}-\ref{eqn:continuum-connection-3}) and Berry curvature $\mathcal{F}_{12}(\vk)$ in Eq.~(\ref{eqn:continuum-curvature}) are defined using the reciprocal coordinates $(k_1, k_2)$. 
In fact, these objects are one-form and two-form in the {\it dual vector space}. 
Hence, these quantities transform differently from vectors in the momentum space. 
In Cartesian coordinates, Berry connection and curvature have to be defined in the following forms:
\begin{align}
    A_x(\vk) &= -iu^\dagger_n(\vk)\frac{\partial}{\partial k_x}u_n(\vk)\,,\\
    A_y(\vk) &= -iu^\dagger_n(\vk)\frac{\partial}{\partial k_y}u_n(\vk)\,,\\
    \mathcal{F}_{xy}(\vk) &= \frac{\partial A_y(\vk)}{\partial k_x} - \frac{\partial A_x(\vk)}{\partial k_y}\,.
\end{align}
When the basis vectors of the Bravais lattice are orthogonal unit vectors, the reciprocal coordinate and the Cartesian coordinate are equivalent. However, when the primitive unit cells are no longer square, for example, in a triangular lattice, these two coordinate frames will lead to noticeable difference.

\subsection{Quantum geometry}\label{sec:qgt}

The Berry curvature $\mathcal{F}_{xy}(\vk)$ is not the only quantity which is responsible for characterizing the Bloch wave functions in Chern bands. 
Another gauge invariant quantity $g_{ij}(\vk)$, dubbed \emph{Fubini-Study metric} (FSM), is also widely studied in topological non-trivial bands. Similar to $\mathcal{F}_{xy}(\vk)$, FSM that corresponds to the $n$-th band is also defined as derivatives of Bloch states \cite{cheng_quantum_2013}:
\begin{align}
    g_{ij}(\vk) =& \frac{1}{2}\sum_{\alpha}\Big{(}\partial_{k_i} u_{\alpha, n}^*(\vk) \partial_{k_j}u_{ \alpha, n}(\vk) + \partial_{k_j} u_{\alpha, n}^*(\vk) \partial_{k_i}u_{\alpha,n}(\vk)\Big{)}\nonumber\\
    & + \sum_{\alpha\beta}u^*_{\alpha, n}(\vk)\partial_{k_i}u_{n,\alpha}(\vk)u^*_{\beta, n}(\vk)\partial_{k_j}u_{\beta,n}(\vk)\,.
\end{align}
Due to the positive definiteness of the FSM, the following inequality always holds:
\begin{equation}\label{eqn:trace-bound}
    {\rm Tr}\,g(\vk) \geq |\mathcal{F}_{xy}(\vk)|\,.
\end{equation}
The (magnitude of) Berry curvature serves the role of the lower bound of FSM. 
Hence, the integral of the FSM over the Brillouin zone will be bounded by the winding number of the band of interest. 

The functional $F[W]$ is related to the quantum geometry of a Bloch band. It is well known that $F[W]$ can be written as integrals of quantities defined in reciprocal space as follows \cite{Marzari1997Maximally}:
\begin{align}
    F[W] =& \Omega_c\int \frac{d^2k}{(2\pi)^2} {\rm Tr}\,g(\vk) + \Omega_c \int \frac{d^2k}{(2\pi)^2} \mathbf{A}^2(\vk) - \left(\Omega_c \int \frac{d^2k}{(2\pi)^2} \mathbf{A}(\vk) \right)^2\,.
\end{align}
Once $F[W]$ is written in this form, the gauge dependency becomes transparent. The first term, which contains the integral of the FSM, is obviously gauge invariant, and the following two terms are both related to the gauge dependent Berry connection.

\section{Gauge choice of Bloch states in 2D}\label{sec:gauge}

In this section, we provide an analytic derivation to obtain the smooth gauge for a Bloch band without using the steepest descent method. We first formulate a smooth gauge with discontinuity solely occurring at the ``boundary'' of the Brillouin zone for 2D Chern bands in Sec.~\ref{sec:smooth-gauge}. Based on this smooth gauge, we outline a procedure for generating a new gauge choice in Sec.~\ref{sec:vortex-gauge}, featuring a vortex at an arbitrarily chosen point within the Brillouin zone, utilizing the Weierstrass function. Finally, we perform one more gauge transformation in Sec.~\ref{sec:coulomb-gauge}, which minimizes the corresponding Wannier function spread defined in Eq.~(\ref{eqn:def-wflf}).

\subsection{Smooth gauge}\label{sec:smooth-gauge}
In this subsection, we follow the method of gauge fixing introduced in Ref.~\cite{soluyanov_smooth_2012}. For a trivial band with $\mathcal{C} = 0$, this method can directly obtain a smooth gauge with continuous Berry connection in the whole Brillouin zone. 
If the band carries a non-vanishing Chern number, this method is also able to find a gauge such that the Berry connection is smooth almost everywhere, with the exception of the ``boundary'' of the Brillouin zone.
In the following, we will assume $\mathcal{C} \neq 0$.

We first choose the gauge of the Bloch state at $(K_1, K_2) = (0, 0)$ randomly, since it will only lead to a global phase rotation in all Bloch states, and it will not affect resulting Berry connection and curvature. Then we choose the phase of the Bloch states $u_n(K_1, K_2=0)$ for $K_1 = 1, 2,\cdots, N_1 - 1$, such that the connection $a_1$ along this axis is uniform:
\begin{equation}
    a_1(K_1, 0) = \frac{\theta_1(0)}{N_1}\,.
\end{equation}
Next, we choose the phase of the Bloch state at $u_n(K_1, K_2)$ with $K_2 \neq 0$ or $N_2 - 1$ to be the same as $u_n(K_1, K_2 - 1)$. More precisely, the gauge of the state $u_n(K_1, K_2)$ is chosen such that the following condition is satisfied:
\begin{equation}
    {\rm Im}\left(u^\dagger_n(K_1, K_2 - 1) u_n(K_1, K_2)\right) = 0\,.
\end{equation}
This gauge choice leads to vanishing $a_2$ component of the Berry connection:
\begin{equation}
    a_2(K_1, K_2) = 0\,,~~ K_2 = 0, 1, \cdots, N_2 - 2\,.
\end{equation}

Since the $a_2$ components are vanishing, we can use the Berry curvature in each plaquette to write the $a_1$ component of the Berry connection away from $K_2 = 0$ line:
\begin{equation}
    a_1(K_1, K_2) = \frac{\theta_1(0)}{N_1} - \sum_{K_2'=0}^{K_2 - 1}\Omega(K_1, K_2')\,.
\end{equation}
Thus, we conclude that $a_1$ component of this gauge choice does not diverge anywhere in the Brillouin zone.

However, the $a_2$ component on the ``boundary'' of the Brillouin zone cannot be chosen arbitrarily. By definition, the connection on the boundary is given by: 
\begin{equation}
    a_2(K_1, N_2 - 1) = {\rm arg}\left(u^\dagger_n(K_1, N_2 - 1) u_n(K_1, N_2)\right)\,,
\end{equation}
and the phase of the Bloch state $u_n(K_1, N_2)$ is already determined by the gauge choice of the state $u_n(K_1, 0)$. In fact, the Berry connection on the boundary can be solved using the definition of the gauge invariant Wilson loop in Eq.~(\ref{eqn:def-wilson-2}):
\begin{equation}\label{eqn:smooth-gauge-jump-1}
    a_2(K_1, N_2 - 1) = \theta_2(K_1)\,.
\end{equation}
Since the Wilson loop exponent $\theta_2(K_1)$ is gauge invariant, it will remain unchanged even if we take the limit $N_1, N_2 \rightarrow \infty$. Thus, the continuum connection component $A_2(k_1,k_2)$ diverges near the line $k_2 = 2\pi$. 

Although the phase discontinuity is confined to a single line, its specific value, as illustrated in Eq.~(\ref{eqn:smooth-gauge-jump-1}), is determined by the Wilson loops, which in turn rely on the detail of the underlying Bloch Hamiltonian. In order to further improve the gauge choice, we perform the following gauge transformation to remove the discontinuity along this line:
\begin{align}
    u_n(K_1, K_2) & \rightarrow e^{i\chi_s(K_1, K_2)} u_n(K_1, K_2)\,,\\
    \chi_s(K_1, K_2) &= \theta_2(K_1)\frac{K_2}{N_2}\,.
\end{align}
Then using the gauge transformation for the Berry connection Eqs.~(\ref{eqn:connection-gauge-1}-\ref{eqn:connection-gauge-2}), we obtain the new expressions for the Berry connection along $\mathbf{b}_2$ axis:
\begin{equation}
    a_2(K_1, K_2) = \frac{\theta_2(K_1)}{N_2}\,.
\end{equation}
Although the value of $a_2$ may still have a jump when $K_1$ goes from $K_1 = N_1 - 1$ back to $K_1 = 0$, it does not have any divergent terms over the Brillouin zone even if $\mathcal{C} \neq 0$. 
In contrast, the $a_1$ components will be transformed into:
\begin{align}
    a_1(K_1, K_2) =& \frac{\theta_1(0)}{N_1} - \sum_{K_2' = 0}^{K_2 - 1}\Omega(K_1, K_2')  + \frac{K_2}{N_2}\left[\theta_2(K_1 + 1) - \theta_2(K_1)\right]\,,\nonumber \\
    &~~\text{if~} K_1 < N_1 - 1;\\
    a_1(N_1 - 1, K_2) =& \frac{\theta_1(0)}{N_1} - \sum_{K_2'=0}^{K_2 - 1}\Omega(N_1 - 1, K_2')  + \frac{K_2}{N_2}\left[\theta_2(N_1) - \theta_2(N_1 - 1)\right] - 2\pi \mathcal{C}\frac{K_2}{N_2}\,;\nonumber \\
    &~~\text{if~} K_1 = N_1 - 1\,.
\end{align}
Here we used the periodicity condition of the Wilson loop Eq.~(\ref{eqn:wilson-period-2}). 
Along the ``boundary'' of the Brillouin zone $k_1 = 2\pi$, the $a_1$ component of the Berry connection has an extra term $-2\pi \mathcal{C}K_2/N_2$, which does not scale as $1/N_1$. 
Consequently, its continuous version, $A_1(k_1=2\pi, k_2)$, will be divergent.
In summary, we have found a gauge that is mostly smooth in the Brillouin zone with the exception along the line $k_1 = 2\pi$. Furthermore, this extra term is independent of the detail of Bloch states, and depends only on the total Chern number of the given band. 

For convenience, we will denote this gauge as $a^s_i$, and its ``continuum'' counterpart as $A_i^s$ or $\mathbf{A}^s$. Using the definitions in Eqs.~(\ref{eqn:continuum-connection-2}-\ref{eqn:continuum-connection-3}), the components of $A_i^s$ can be written as:
\begin{align}
    A^s_1(k_1, k_2) =& \frac{\theta_1(0)}{2\pi} - \int_0^{k_2}dk_2'~ \mathcal{F}_{12}(k_1,k_2') + \frac{k_2}{2\pi}\frac{d\theta_2(k_1)}{dk_1} - \mathcal{C}k_2\delta(k_1 - 2\pi)\,.\\
    A^s_2(k_1, k_2) =& \frac{\theta_2(k_1)}{2\pi}\,.
\end{align}
The derivatives of the Berry connection can also be computed as follows:
\begin{align}
    \frac{\partial A^s_1}{\partial k_1} =& -\int_0^{k_2}dk_2'~\frac{\partial}{\partial k_1}\mathcal{F}_{12}(k_1, k_2') + \frac{k_2}{2\pi}\frac{d^2\theta_2(k_1)}{dk_1^2} - \mathcal{C}k_2 \frac{d \delta(k_1 - 2\pi)}{d k_1}\,,\label{eqn:connection-derivative1}\\
    \frac{\partial A^s_2}{\partial k_2} =& 0\,,\label{eqn:connection-derivative2}\\
    \frac{\partial A^s_1}{\partial k_2} =& -\mathcal{F}_{12}(k_1, k_2) + \frac{1}{2\pi}\frac{d\theta_2(k_1)}{dk_1} - \mathcal{C}\delta(k_1 - 2\pi)\,,\label{eqn:connection-derivative3}\\
    \frac{\partial A^s_2}{\partial k_1} =& \frac{1}{2\pi}\frac{d \theta_2(k_1)}{d k_1} - \mathcal{C}\delta(k_1 - 2\pi)\,.\label{eqn:connection-derivative4}
\end{align}
Here the $\delta$ function in Eq.~(\ref{eqn:connection-derivative4}) originates from both the single-valued nature of $A_2^s$ and the winding behavior of the Wilson loop exponent. These expressions will be useful when we discuss the Coulomb gauge condition in Sec.~\ref{sec:coulomb-gauge}.

\subsection{Weierstrass \texorpdfstring{$\sigma$}{sigma} function and vortex gauge transformations}\label{sec:vortex-gauge}

The smooth gauge Berry connection $a_i^s$ obtained in the previous subsection has a phase jump along the line $k_1 = 2\pi$ if there is a topological obstruction. In this subsection, we discuss how this linear-discontinuity can be replaced by a vortex at an arbitrary position in the Brillouin zone. 

Assuming the existence of a gauge transformation $e^{i\chi_v(\vk)}$ capable of eliminating the linear discontinuity and substituting it with a vortex at $\vk_v$, the following conditions must be met:
\begin{enumerate}
    \item The change of $\chi_v(\vk) \rightarrow \chi_v(\vk + \mathbf{b}_1)$ has to compensate the phase jump $\sim -\mathcal{C} k_2$ at $k_1 = 2\pi$: 
    \begin{equation*}
        \chi_v(k_1 + 2\pi, k_2) - \chi_v(k_1, k_2) ~{\rm mod}~2\pi = -\mathcal{C}k_2\,.
    \end{equation*}
    
    \item Since the smooth gauge $a^s_i$ is already continuous along $\mathbf{b}_2$ direction, $e^{i\chi_v(\vk)}$ is periodic under the momentum translation $\vk \rightarrow \vk + \mathbf{b}_2$:
    \begin{equation*}
        \chi_v(k_1, k_2 + 2\pi) - \chi_v(k_1, k_2)~{\rm mod}~2\pi = 0\,.
    \end{equation*}

    \item The transformation function $e^{i\chi_v(\vk)}$ is continuous in the Brillouin zone except for one point $\vk_v$. The value of $\chi_v(\vk)$ increase by $2\pi \mathcal{C}$ as $\vk$ undergoes a clockwise rotation around $\vk_v$.
\end{enumerate}

To solve the gauge transformation which satisfies all these conditions, we first map each point $\vk$ in the momentum space to a complex number as $k = k_x + i k_y$, and map each reciprocal vector to complex numbers $G = G_1 \omega_1 + G_2 \omega_2$, with $G_1, G_2\in \mathbb{Z}$, $\omega_1 = \hat{\mathbf{x}}\cdot\mathbf{b}_1 + i \hat{\mathbf{y}}\cdot\mathbf{b}_1$ and $\omega_2 = \hat{\mathbf{x}}\cdot\mathbf{b}_2 + i \hat{\mathbf{y}}\cdot\mathbf{b}_2$. Therefore, the first and second quasi-periodic conditions can be written as the following form using the complex number representation:
\begin{align}
    \chi_v(k + \omega_1) - \chi_v(k)~{\rm mod}~2\pi &= -2\pi\mathcal{C}\frac{k\omega_1^* - k^* \omega_1}{\omega_2\omega_1^* - \omega_2^*\omega_1}\,,\label{eqn:complex-period-1}\\
    \chi_v(k + \omega_2) - \chi_v(k)~{\rm mod}~2\pi &= 0\,,\label{eqn:complex-period-2}
\end{align}
in which the fraction on the right hand side of Eq.~(\ref{eqn:complex-period-1}) is the complex number representation of $k_2$.

Surprisingly, these conditions can be fulfilled with a double-quasi-periodic entire function, the Weierstrass sigma function \cite{whittaker1920course, gunawardana2023optimally}, whose definition is given by:
\begin{equation}
    \sigma(k; \mathcal{G}) = k \prod_{G \in \mathcal{G}\text{\textbackslash}\{0\}} \left(1 - \frac{k}{G}\right)\exp\left(\frac{k}{G} + \frac{k^2}{2G^2}\right)\,,
\end{equation}
in which $\mathcal{G} = \left\{G| G = G_1 \omega_1 + G_2 \omega_2; G_1, G_2 \in \mathbb{Z}\right\}$ is the set of the complex number reciprocal lattice sites, with $\omega_1$ and $\omega_2$ being the two periods, which correspond to the two reciprocal basis vectors.
The asymptotic behavior of this function around $k\rightarrow 0$ is given by:
\begin{equation}
    \lim_{k\rightarrow 0} \frac{\sigma(k; \mathcal{G})}{k} = 1\,.
\end{equation}
Clearly, the imaginary part of $\log \sigma(k; \mathcal{G})$ will have a vortex around the origin. This indicates that the third condition might be satisfied by choosing $\chi_v$ as the imaginary part of $\log \sigma(k; \mathcal{G})$ with proper coefficients.

The Weierstrass sigma function also satisfies the following quasi-periodic conditions:
\begin{align}
    \sigma(k + \omega_1; \mathcal{G}) &= -\exp\left[\eta_1\left(k + \frac{\omega_1}{2}\right)\right]\sigma(k; \mathcal{G})\,,\\
    \sigma(k + \omega_2; \mathcal{G}) &= -\exp\left[\eta_2\left(k + \frac{\omega_2}{2}\right)\right]\sigma(k; \mathcal{G})\,,
\end{align}
in which $\eta_1$ and $\eta_2$ are the parameters of $\sigma(k;\mathcal{G})$ and are uniquely determined by the values of $\omega_1, \omega_2$ as explained in Sec.~\ref{appsec:Elliptic-functions}. They also satisfy the identity $\omega_2 \eta_1 - \omega_1 \eta_2 = 2\pi i$ \cite{whittaker1920course}. Taking the logarithmic function on both sides of the quasi-periodic condition, we find:
\begin{align}
    \log \sigma(k + \omega_1; \mathcal{G}) &= \log \sigma(k; \mathcal{G}) + \eta_1\left(k + \frac{\omega_1}{2}\right) + i\pi\,,\label{eqn:sigma-period-1}\\
    \log \sigma(k + \omega_2; \mathcal{G}) &= \log \sigma(k; \mathcal{G}) + \eta_2\left(k + \frac{\omega_2}{2}\right) + i\pi\,,\label{eqn:sigma-period-2}
\end{align}
in which a linear term of $k$ can be generated when $k\rightarrow k + \omega_i$. Thus, the first and second conditions might also be satisfied with the $\log\sigma(k; \mathcal{G})$ function if proper parameters are used.

With the aforementioned properties of the Weierstrass sigma function, we assume the vortex gauge transformation $\chi_v(k)$ has the following form:
\begin{equation}\label{eqn:vortex-gauge-transformation-solution}
    \chi_v(k) = -\mathcal{C}~\mathrm{Im}\left(\log \sigma(k - k_v; \mathcal{G}) + A k^2 + Bk^* k + Ck\right)\,,
\end{equation}
in which $k_v = \hat{\mathbf{x}}\cdot \vk_v + i \hat{\mathbf{y}}\cdot \vk_v$ is the complex number representation of the vortex location in the momentum space. 
By matching the quasi-periodic boundary conditions in Eqs.~(\ref{eqn:complex-period-1}-\ref{eqn:complex-period-2}) and Eqs.~(\ref{eqn:sigma-period-1}-\ref{eqn:sigma-period-2}), the values of the parameters $A$, $B$ and $C$ can be solved directly:
\begin{align}
    B &= \frac{\pi i (\omega_1^* \omega_2 + \omega_2^*\omega_1)}{(\omega_2 \omega_1^* - \omega_2^* \omega_1)^2} = -\frac{\pi i \mathbf{b}_1\cdot\mathbf{b}_2}{2V_{\rm BZ}^2}\,,\\
    A &= -\frac{2 B \omega_2^* + \eta_2}{2\omega_2}\,,\\
    C &= \frac{2i[\omega_2^*(\pi - \xi_1) - \omega_1^*(\pi - \xi_2)]}{\omega_1 \omega_2^* - \omega_2 \omega_1^*}\,.
\end{align}
Here $V_{\rm BZ} = 4\pi^2/\Omega_c = {\rm Im}(\omega_2\omega_1^* - \omega_2^*\omega_1)$ is the volume of the Brillouin zone, and the quantities $\xi_1, \xi_2$ are given by:
\begin{equation}
    \xi_i = {\rm Im}\left(\frac{\eta_i\omega_i}{2} - \eta_i k_v + A \omega_i^2 + B \omega_i^*\omega_i\right)\,.
\end{equation}

In Fig.~\ref{fig:vortex-gauge-transformation}, we provide an example of the vortex gauge transformation in the Brillouin zone of a square lattice with $\omega_1 = 2\pi, \omega_2 = 2\pi i$, Chern number $\mathcal{C} = 1$ and vortex at $\vk_v = \frac12 (\mathbf{b}_1 + \mathbf{b}_2)$.

Applying the vortex gauge transformation $\chi_v(\vk)$ to Bloch states with the smooth gauge that obtained in Sec.~\ref{sec:vortex-gauge}, the discontinuity of the connection along $k_1 = 2\pi$ will be removed, and a vortex at $\vk_v$ will also be generated. We denote the Berry connection after the vortex gauge transformation as $a_i^v(K_1, K_2)$ or $\mathbf{A}^v(\vk)$.

\begin{figure}
    \centering
    \includegraphics[width=0.5\linewidth]{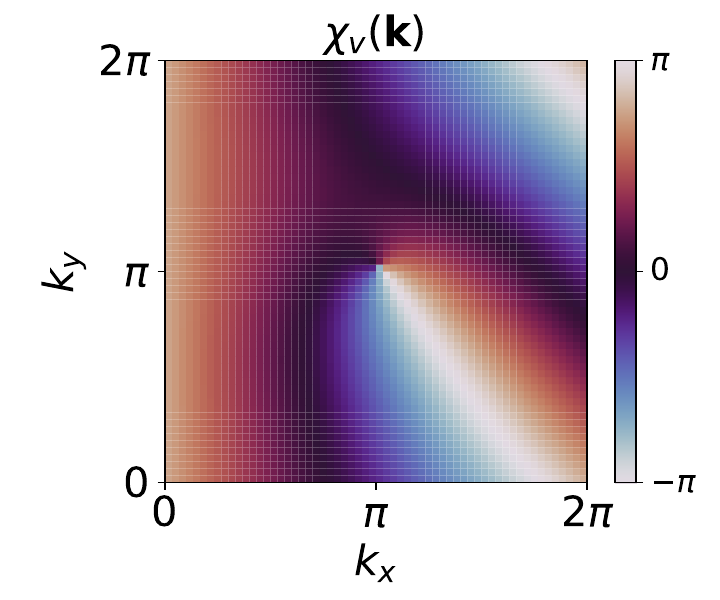}
    \caption{An example of a vortex gauge transformation $\chi_v(\vk)$ evaluated using Eq.~(\ref{eqn:vortex-gauge-transformation-solution}). Here we use the Brillouin zone of a 2D square lattice, and we assume the Chern number $\mathcal{C} = 1$, with the vortex located at $\vk_v = \frac{1}{2}(\mathbf{b}_1 + \mathbf{b}_2)$.}
    \label{fig:vortex-gauge-transformation}
\end{figure}

\subsection{Coulomb gauge and Poisson's equation}\label{sec:coulomb-gauge}
By performing the functional derivative of the WFLF with respect to the smooth gauge choices of the Bloch states, we can establish the condition for achieving maximal Wannier function localization.
This condition is met when the Bloch states satisfy the Coulomb gauge condition:
\begin{equation}\label{eqn:coulomb-gauge}
    \nabla_\vk \cdot \mathbf{A}(\vk) \equiv \partial_{k_x}A_x + \partial_{k_y}A_y = 0\,.
\end{equation}
Here, the divergence is defined using {\it Cartesian coordinates} \cite{blount_formalisms_1962}, which originates from the Euclidean distance in the definition of the functional $F[W]$.

In order to calculate the divergence of the smooth gauge Berry connection, we have to write the components of both $\vk$ and $\mathbf{A}(\vk)$ using the Cartesian basis as:
\begin{align}
    \left(
        \begin{array}{c}
           k_x\\
           k_y
        \end{array}
    \right) &= \frac{1}{2\pi} \left(
        \begin{array}{cc}
            \mathbf{b}_1\cdot \hat{\mathbf{x}} & \mathbf{b}_2\cdot \hat{\mathbf{x}} \\
            \mathbf{b}_1\cdot \hat{\mathbf{y}} & \mathbf{b}_2\cdot \hat{\mathbf{y}}
        \end{array}
    \right) \left(
        \begin{array}{c}
           k_1\\
           k_2
        \end{array}
    \right)\,,\\
    \left(
        \begin{array}{c}
           A_1\\
           A_2
        \end{array}
    \right) &= \frac{1}{2\pi} \left(
        \begin{array}{cc}
            \mathbf{b}_1\cdot \hat{\mathbf{x}} & \mathbf{b}_1\cdot \hat{\mathbf{y}} \\
            \mathbf{b}_2\cdot \hat{\mathbf{x}} & \mathbf{b}_2\cdot \hat{\mathbf{y}}
        \end{array}
    \right) \left(
        \begin{array}{c}
           A_x\\
           A_y
        \end{array}
    \right)\,.
\end{align}
Consequently, the partial derivative operators and the Berry connection vectors under the Cartesian basis are given by:
\begin{align}
    \frac{\partial}{\partial k_x} &= \frac{2\pi}{V_{\rm BZ}}\left(\mathbf{b}_2\cdot\hat{\mathbf{y}}\frac{\partial}{\partial k_1} - \mathbf{b}_1 \cdot \hat{\mathbf{y}}\frac{\partial}{\partial k_2}\right)\,,\\
    \frac{\partial}{\partial k_y} &= \frac{2\pi}{V_{\rm BZ}}\left(-\mathbf{b}_2\cdot\hat{\mathbf{x}}\frac{\partial}{\partial k_1} + \mathbf{b}_1 \cdot \hat{\mathbf{x}}\frac{\partial}{\partial k_2}\right)\,,\\
    A_x &= \frac{2\pi }{V_{\rm BZ}}\left(\mathbf{b}_2\cdot\hat{\mathbf{y}} A_1 - \mathbf{b}_1\cdot \hat{\mathbf{y}} A_2\right)\,,\\
    A_y &= \frac{2\pi }{V_{\rm BZ}}\left(-\mathbf{b}_2\cdot\hat{\mathbf{x}} A_1 + \mathbf{b}_1\cdot \hat{\mathbf{x}} A_2\right)\,.
\end{align}
Using these identities, the divergence of the connection will become the following form:
\begin{align}
    \nabla_\vk \cdot \mathbf{A}(\vk) &= \frac{\partial A_x}{\partial k_x} + \frac{\partial A_y}{\partial k_y} \nonumber\\
    &= \frac{4\pi^2}{V_{\rm BZ}^2}\left[|\mathbf{b}_2|^2\frac{\partial A_1}{\partial k_1} + |\mathbf{b}_1|^2\frac{\partial A_2}{\partial k_2} - \mathbf{b}_1\cdot \mathbf{b}_2\left(\frac{\partial A_2}{\partial k_1} + \frac{\partial A_1}{\partial k_2}\right)\right]\,.\label{eqn:div-cartesian-to-b1b2}
\end{align}
The Laplacian of a scalar function $\chi(\vk)$ has a similar expression:
\begin{align}
    \nabla_\vk^2 \chi(\vk) &= \frac{4\pi^2}{V_{\rm BZ}^2}\left[|\mathbf{b}_2|^2\frac{\partial^2 \chi}{\partial k_1^2} + |\mathbf{b}_1|^2\frac{\partial^2 \chi}{\partial k_2^2} - 2\mathbf{b}_1\cdot \mathbf{b}_2\frac{\partial^2 \chi}{\partial k_1\partial k_2}\right]\,.\label{eqn:laplacian-cartesian-to-b1b2}
\end{align}

Next, we will check whether the gauge choice we obtained in Sec.~\ref{sec:vortex-gauge} with the vortex $A_i^v(k_1, k_2)$ satisfies the Coulomb gauge condition. Since the vortex gauge is obtained by the transformation $e^{i\chi_v(\vk)}$ from the smooth gauge $\mathbf{A}^s(\vk)$, the divergence of the Berry connection with vortex gauge can be written as:
\begin{align}
    \nabla_\vk \cdot \mathbf{A}^v(\vk) &= \nabla_\vk \cdot \mathbf{A}^s(\vk) + \nabla^2_\vk \chi_v(\vk)\,.
\end{align}
Note that all the terms in the expression of $\chi_v(\vk)$ except for $B k^* k$ are the imaginary parts of holomorphic functions, we conclude that the Laplacian of the transformation $\chi_v(\vk)$ will only contain the contribution from the $B k^* k$ term if $k_1 \neq 2\pi$,  because of the Cauchy-Riemann equation:
\begin{equation}
    \nabla^2_\vk \chi_v(\vk) = -4\mathcal{C}~{\rm Im\,}B = \frac{2\pi \mathcal{C} \mathbf{b}_1\cdot \mathbf{b}_2}{V_{\rm BZ}^2}\,.
\end{equation}
Moreover, the vortex gauge transformation $\chi_v(\vk)$ also has value jumps near the divergent line $k_1 = 2\pi$. This will lead to an extra $\delta$ function contribution to the derivative of the transformation:
$$
\frac{\partial \chi_v}{\partial k_1} \sim \mathcal{C}k_2 \delta(k_1 - 2\pi)\,.
$$ 
Thus, the Laplacian of the vortex gauge transformation is:
\begin{align}
     \nabla^2_\vk \chi_v(\vk) =& \frac{2\pi \mathcal{C} \mathbf{b}_1\cdot \mathbf{b}_2}{V_{\rm BZ}^2}  + \frac{4\pi^2}{V^2_{\rm BZ}}\mathcal{C} \Bigg{[} |\mathbf{b}_2|^2k_2\frac{d \delta(k_1 - 2\pi)}{d k_1} - 2\mathbf{b}_1\cdot \mathbf{b}_2\delta(k_1 - 2\pi) \Bigg{]}\,.\label{eqn:vortex-laplacian-w-boundary}
\end{align}

Combining Eq.~(\ref{eqn:div-cartesian-to-b1b2}) with the expression of connection derivatives in Eqs.~(\ref{eqn:connection-derivative1}-\ref{eqn:connection-derivative4}), and the expression for the Laplacian of the vortex gauge transformation Eq.~(\ref{eqn:vortex-laplacian-w-boundary}), we found that the discontinuity terms cancel with each other.
Eventually, we obtain the expression for $\nabla_\vk\cdot \mathbf{A}^v(\vk)$ as shown:
\begin{equation}\label{eqn:div-A-vortex}
    \nabla_\vk \cdot \mathbf{A}^v(\vk)= \frac{4\pi^2}{V^2_{\rm BZ}}\Bigg{[}|\mathbf{b}_2|^2\left(-\int_0^{k_2}dk_2'\frac{\partial}{\partial k_1}\mathcal{F}_{12}(k_1, k_2') + \frac{k_2}{2\pi}\frac{d^2\theta_2(k_1)}{dk_1^2}\right) - \mathbf{b}_1\cdot\mathbf{b}_2\left(-\mathcal{F}_{12}(k_1, k_2) + \frac{1}{\pi}\frac{d\theta_2(k_1)}{dk_1} - \frac{\mathcal{C}}{2\pi}\right) \Bigg{]}\,.
\end{equation}
Although the Berry connection itself $\mathbf{A}^v(\vk)$ has a singularity, its divergence remains smooth over the whole Brillouin zone. In general, $\nabla_\vk\cdot\mathbf{A}^v(\vk)$ is not zero as shown in Eq.~(\ref{eqn:div-A-vortex}), thus $\mathbf{A}^v(\vk)$ does not satisfy the Coulomb gauge condition. To find a gauge which satisfies this condition, we proceed with an additional gauge transformation, denoted as $\chi_c(\vk)$, applied to $\mathbf{A}^v(\vk)$. The Coulomb gauge condition can indeed be rewritten as a Poisson's equation:
\begin{equation}
    \nabla^2_\vk \chi_c(\vk) = -\nabla_\vk \cdot \mathbf{A}^v(\vk)\,,
\end{equation}
in which the right hand side of the equation is a known function given by Eq.~(\ref{eqn:div-A-vortex}). One can easily check that the integral of $\nabla_\vk \cdot \mathbf{A}^v(\vk)$ over the whole Brillouin zone is zero. 
This condition allows a smooth solution for $\chi_c(\vk)$ \cite{donaldson2011riemann}.
Hence, this Poisson's equation can be solved by Fourier transformation:
\begin{align}
    \chi_c(\vk) &= \sum_{\mathbf{R}\neq 0}\frac{\mathcal{A}_\mathbf{R}}{|\mathbf{R}|^2}e^{i\vk\cdot \mathbf{R}} + {\rm const.}\,,\\
    \mathcal{A}_\mathbf{R} &= \frac{1}{V_{\rm BZ}}\int_{\rm BZ}dk_xdk_y \nabla_\vk\cdot\mathbf{A}^v(\vk)e^{-i\vk\cdot \mathbf{R}}\,.
\end{align}
The Berry connection will have a vanishing divergence after this gauge transformation.

Following the procedure described in Secs.~\ref{sec:smooth-gauge}, \ref{sec:vortex-gauge} and \ref{sec:coulomb-gauge}, we eventually yield a proper gauge choice of the Bloch states which is smooth over the Brillouin zone with an exception of a vortex point, while at the same time satisfies the Coulomb gauge condition. 
We note that this procedure can be easily applied to both tight-binding Bloch states and continuum Bloch states, and this procedure can also be easily generalized to the case with multiple vortices, as long as the total vorticity is equal to the Chern number.

\subsection{Position of the vortex}\label{sec:position-gauge}

The Coulomb gauge condition is derived through variations of the WFLF over smooth gauge transformations. However, the position of the vortex is the other degree of freedom in the gauge choice. One of the major conclusion of Ref.~\cite{gunawardana2023optimally} is the dependency of the functional $F[W]$ on the value of $\vk_v$. If the gauge choice of the Bloch states contains a vortex at $\vk_v$, $F[W]$ will have the following form:
\begin{align}
    F[W] &\sim {\rm const.} - \frac{4\pi\mathcal{C}}{V_{\rm BZ}}\phi(\vk_v)\,,\label{eqn:vortex-position-wflf}\\
    \nabla^2_\vk \phi(\vk) &= \frac{2\pi\mathcal{C}}{V_{\rm BZ}} - \mathcal{F}_{xy}(\vk)\,,\label{eqn:vortex-position-poisson}
\end{align}
in which the constant term in $F[W]$ is actually divergent for bands with $\mathcal{C} \neq 0$, and is independent of the position of the vortex. The function $\phi(\vk)$ can be solved from the Poisson's equation with the Berry curvature $\mathcal{F}_{xy}(\vk)$ being the inhomogeneous term. Thus, we are able to obtain the optimally localized Wannier function of a given Bloch band by placing the vortex at the minimum (maximum) position of the function $\phi(\vk)$. We denote this function $\phi(\vk)$ as \emph{smooth vortex potential}, since it serves the role of the smooth ``potential energy'' experienced by the vortex in the electrostatics analogy of Berry connection, as introduced in Ref.~\cite{gunawardana2023optimally}. We note that the extreme value of the smooth vortex potential $\phi(\vk)$ might \emph{not} locate at high symmetry points. Thus, the \emph{most} optimally localized Wannier function may break more symmetries than Wannier function gauge choices exhibiting larger values of $F[W]$, as we will see in Sec.~\ref{sec:models}.

The position of the vortex in the Brillouin zone also affects the ``charge center'' $\mathbf{R}_c = \langle W_n(\mathbf{0}) | \hat{\mathbf{r}} | W_n(\mathbf{0})\rangle$. Due to the winding of the Wilson loop phases $\theta_1(k_2), \theta_2(k_1)$, the value of Zak phase changes with the choice of the parallelogram Brillouin zone \cite{Coh2009Electric}. As a consequence, the charge center coordinate in real space will change according to the vortex position in the reciprocal space as follows \cite{Coh2009Electric, gunawardana2023optimally}:
\begin{equation}
    \mathbf{R}_c^{(1)} - \mathbf{R}_c^{(2)} = \frac{2\pi\mathcal{C}}{V_{\rm BZ}}\left[\vk_v^{(1)} - \vk_v^{(2)}\right]\times \hat{\mathbf{z}}\,,
\end{equation}
in which $\mathbf{R}_c^{(1,2)}$ and $\mathbf{k}_v^{(1,2)}$ stand for the charge center positions and vortex positions which correspond to two gauge choices of the same Bloch band. Empirically, the relationship between the ``charge center'' and the vortex position shares a notable similarity to the magnetic translation algebra in generalized lowest Landau levels \cite{wang2023origin}.

\subsection{Other gauge transformations}\label{sec:other-gauge}

Using the procedure described in the previous subsections, we have guaranteed that the gauge choice $\mathbf{A}^c(\vk)$ satisfy the Coulomb gauge condition, with a vortex located at $\vk_v$. 
One may still wonder if the gauge choice is {\it uniquely} determined by this procedure.
Let us assume that a transformation $\chi'(\vk)$ can lead to another gauge choice $\mathbf{A}'(\vk) = \mathbf{A}^c(\vk) + \nabla_\vk \chi'(\vk)$, which also satisfies $\nabla_\vk\cdot \mathbf{A}'(\vk) = 0$ with the vortex unchanged. 
Indeed, such gauge transformation has to satisfy the following conditions:
\begin{itemize}
    \item has no singularity over the first Brillouin zone;
    \item and satisfies the Laplace's equation $\nabla^2_\vk \chi'(\vk) = 0$.
\end{itemize}
The second requirement ensures the Coulomb gauge will not be destroyed by $\chi'(\vk)$. 
Generically, such gauge choice can always be written in the following form:
\begin{equation}
    \chi'(\vk) = \vk \cdot \mathbf{R}' + \chi_0(\vk)\,,~~\mathbf{R}' = R_1' \mathbf{a}_1 + R_2' \mathbf{a}_2\,,~~R_1', R_2' \in \mathbb{Z}\,,
\end{equation}
in which the first term is a linear large gauge transformation, and $\chi_0(\vk)$ is a double-periodic function satisfying the Laplace's equation.
Due to Cauchy-Riemann equation, such function can be written as the imaginary part of an analytic function $\chi_0 = {\rm Im}\,g(k_x+ik_y)$. However, we also required that this function is double-periodic and non-singular at the same time, indicating that it is bounded over the whole complex plane. As per Liouville's theorem \cite{whittaker1920course}, $\chi_0$ can only be a constant function.
Thus, the only non-trivial gauge transformations that satisfies these two conditions are linear large gauge transformations. Correspondingly, the Berry connection will be shifted by a constant vector:
\begin{equation}
    \mathbf{A}(\vk)\rightarrow \mathbf{A}(\vk) + \mathbf{R}'\,.
\end{equation}
Obviously, such gauge transformation can shift the Wannier states in real space $w^\dagger_{\mathbf{R},n}\rightarrow w^\dagger_{\mathbf{R} + \mathbf{R}',n}$, but will not affect the ``shape'' of these wave functions. Hence, these gauge transformations are all ``trivial''. We conclude that the Wannier functions that we obtained in Secs.~\ref{sec:smooth-gauge}, \ref{sec:vortex-gauge} and \ref{sec:coulomb-gauge} are uniquely determined, assuming that the vortex location is not changed in the gauge choice.

\section{Models}\label{sec:models}

In this section, we discuss the Wannier functions of two example models. 
We also provide a simple discussion about the gauge choice of the ideal Chern bands, which saturates the FSM inequality, in Sec.~\ref{sec:ideal-chern-band}.

\subsection{Kagome lattice with complex hoppings}\label{sec:kagome}

\begin{figure}[t]
    \centering
    \includegraphics[width=0.75\linewidth]{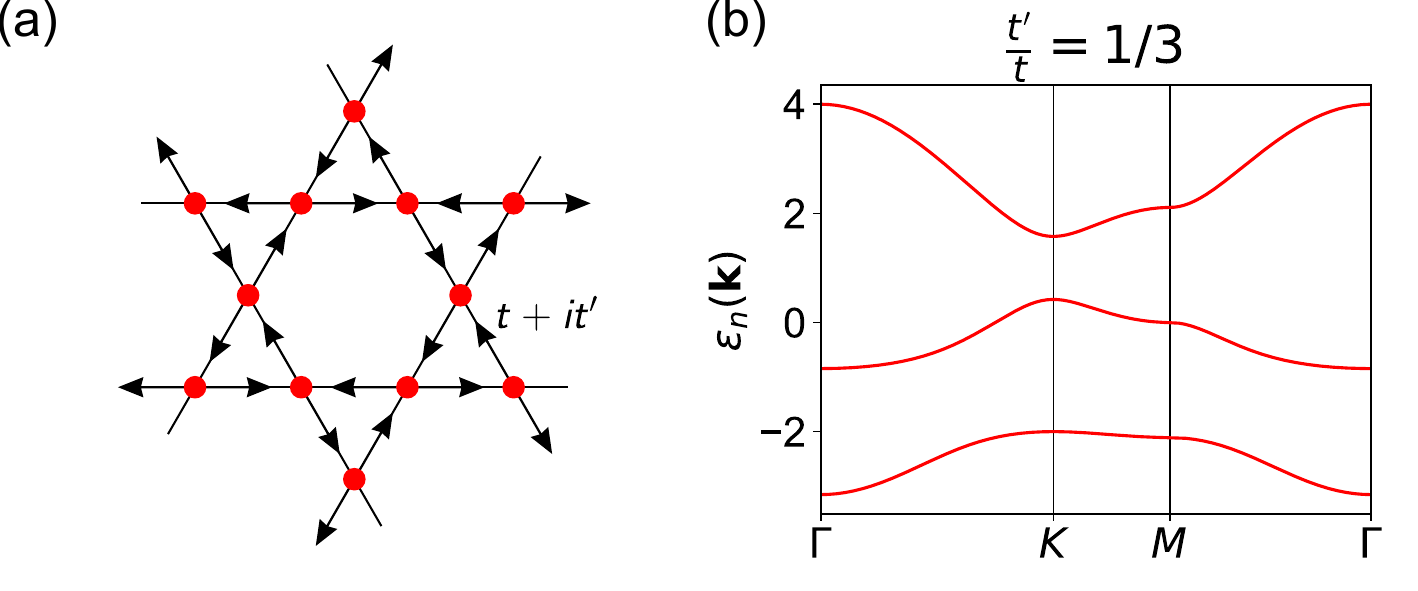}
    \caption{(a) The lattice structure of the kagome model. Hoppings along the arrows are given by $t + i t'$.
    (b) Band structure of the kagome lattice model with $t = 1$ and $t' = 1/3$. The lowest band with a narrow band width carries Chern number $\mathcal{C} = 1$.}
    \label{fig:kagome-bands}
\end{figure}

\begin{figure}[t]
    \centering
    \includegraphics[width=0.5\linewidth]{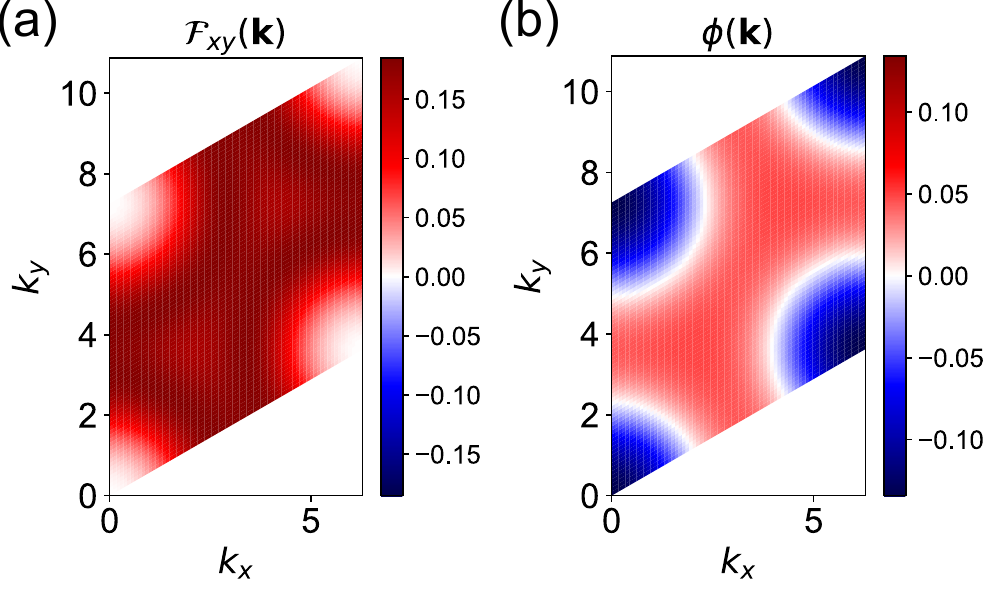}
    \caption{(a) The Berry curvature $\mathcal{F}_{xy}(\vk)$ of the lowest Chern band in the kagome model. (b) The smooth vortex potential of this Chern band. Here we set the hopping parameters to be the same as in Fig.~\ref{fig:kagome-bands}(b).}
    \label{fig:kagome-curvature}
\end{figure}

\begin{figure}[t]
    \centering
    \includegraphics[width=0.75\linewidth]{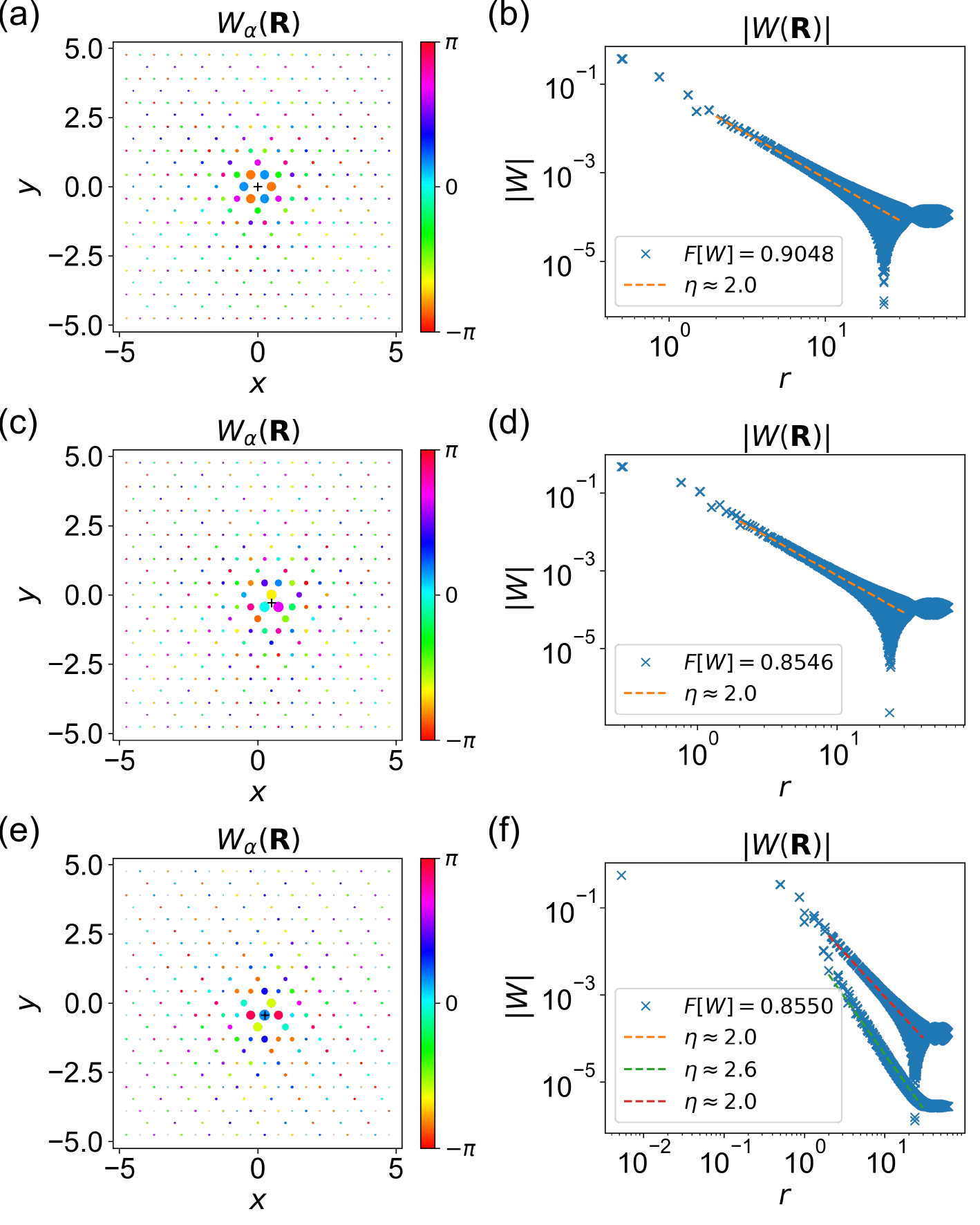}
    \caption{(a-b) Wannier function of the lowest Chern band in the kagome model with the vortex located at $\Gamma$ point. 
    (c-d) Wannier function with the vortex located at $K = (\mathbf{b}_1 + \mathbf{b}_2)/3$ point. 
    (e-f) Wannier function with the vortex located at $M = \mathbf{b}_1/2$ point. 
    The amplitude of the Wannier functions are fitted as $W(\mathbf{R}) \sim 1/R^\eta$.
    The WFLF values are computed on a $72\times 72$ mesh.}
    \label{fig:kagome-wannier}
\end{figure}

\begin{figure}[t]
    \centering
    \includegraphics[width=0.5\linewidth]{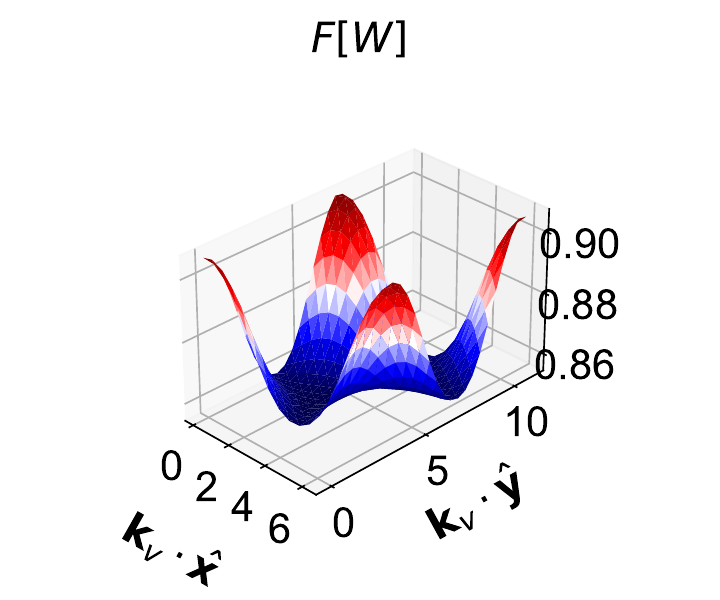}
    \caption{The value of $F[W]$ with different vortex position $\vk_v$. The Bloch states always satisfy the Coulomb gauge condition.}
    \label{fig:kagome-wflf}
\end{figure}

\begin{figure}
    \centering
    \includegraphics[width=0.75\linewidth]{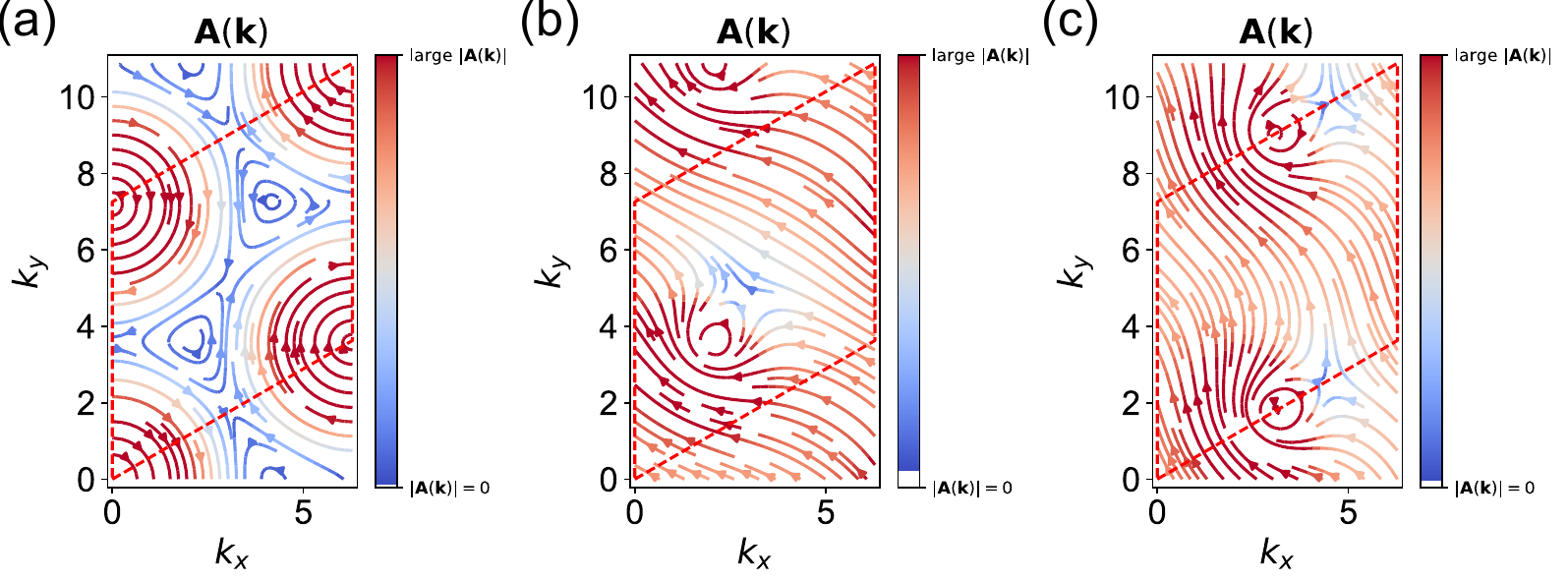}
    \caption{The Berry connection of the lowest Chern band in the kagome tight binding model. In all three cases, the Coulomb gauge condition is satisfied. (a) The vortex is at the $\Gamma$ point. (b) The vortex is at the $K$ point. (c) The vortex is at one of the $M$ points.}
    \label{fig:kagome-connection}
\end{figure}

We choose a three band spinless tight-binding model without time-reversal symmetry on the kagome lattice, which possess a narrow Chern band. It was also once considered to be a host of fractional Chern insulator states \cite{Tang2011High, wu2012zoology}. Besides, with a complementary kagome dispersive band with different inversion symmetry eigenvalue, a pair of molecular orbital can also form in the kagome lattice models, enabling the possibility of Kondo-driven physics in these topological bands \cite{chen2023Metallic}.

The kagome lattice is shown in Fig.~\ref{fig:kagome-bands}(a), and the Bravais lattice basis vectors are $\mathbf{a}_1 = (1, 0)$, $\mathbf{a}_2 = (-1/2, \sqrt{3}/2)$. There are three sublattices within each unit cell, which are located at $\bm{\tau}_1 = \mathbf{a}_1/2$, $\bm{\tau}_2 = \mathbf{a}_2/2$ and $\bm{\tau}_3 = (\mathbf{a}_1 + \mathbf{a}_2)/2$. With the broken time reversal symmetry, the hoppings can have imaginary parts. We assume the hoppings along the directions labeled by the arrows are given by $t + it'$, in which $t, t' \in \mathbb{R}$. The Bloch Hamiltonian can be written as the following form:
\begin{align}
    h(\vk) =& 2t \left( \lambda_6 \cos \kappa_1 + \lambda_4 \cos \kappa_2 + \lambda_1 \cos \kappa_3 \right) - 2 t'\left(\lambda_7 \cos \kappa_1 - \lambda_5 \cos \kappa_2 + \lambda_2 \cos \kappa_3 \right)\,,
\end{align}
in which $\kappa_j = \vk \cdot \bm{\tau}_j$. $\lambda_{6,4,1}$ and $\lambda_{7,5,2}$ are the off-diagonal real and imaginary Gell-Mann matrices, respectively.

The band structure of this model with $t = 1$ and $t' = 1/3$ has been provided in Fig.~\ref{fig:kagome-bands}(b). When the time-reversal breaking term is turned on ($t' \neq 0$), this model no longer possesses $C_{2x}$ symmetry. 
Consequently, the bands cease to be degenerate at $\Gamma$ and $K$ points, presenting a distinct contrast from the time-reversal symmetric case. 
Using a $72\times 72$ momentum mesh in the Brillouin zone, the Berry curvature of the lowest energy band is shown in Fig.~\ref{fig:kagome-curvature}(a), which gives us a non-vanishing Chern number $\mathcal{C} = 1$. 
The smooth vortex potential $\phi(\vk)$ has also been solved from the Berry curvature, which is shown in Fig.~\ref{fig:kagome-curvature}(b). 
Note that the maximum value of $\phi(\vk)$ is around the $K$ and $K'$ points, while the minimum is at the $\Gamma$ point. 

Wannier function with the vortex located at the $\Gamma$ is shown in Figs.~\ref{fig:kagome-wannier}(a-b). 
The wave function decays as $\sim 1/R^2$ on all three sublattices, and its charge center $\mathbf{R}_c$ sits at the $1a$ position. 
Notably, this Wannier function resembles the {\it compact localized states} of the kagome flat band without SOC \cite{rhim_singular_2021}, and the {\it compact molecular orbitals} in multi-orbital kagome lattice models \cite{chen2023Metallic}.
If the vortex is at the $K$ point, as shown in Figs.~\ref{fig:kagome-wannier}(c-d), the center $\mathbf{R}_c$ will be at the $2b$ (hexagon) position. 
In this case, the wave function on all three types of sublattices also decays as $\sim 1/R^2$.
Finally, we place the vortex at one of the $M$ point, and the Wannier function is shown in Figs.~\ref{fig:kagome-wannier}(e-f).
Wave function on one of the three sublattices decays as $\sim 1/R^{2.6}$, while it still decays slower than $1/R^2$ on other two sublattices.
Among these three Wannier functions, the one with the vortex at the $K$ point has the smallest value of $F[W]$, as one would expect from the smooth vortex potential $\phi(\vk)$.
Although the Wannier function with vortex at $M$ point has components which decay faster than $1/R^2$, the value of $F[W]$ is still larger than the Wannier function with vortex located at $K$ point, whose sublattice components all decay as $1/R^2$. 
Therefore, we conclude that the existence of fast-decaying components in the Wannier function does {\it not} imply a smaller $F[W]$ value.
In Fig.~\ref{fig:kagome-wflf}, we also provide the value of $F[W]$ as a function of vortex position, in which $\vk_v$ is chosen on $18\times 18$ points among the $72\times72$ momentum mesh. It indeed follows the smooth vortex potential function shown in Fig.~\ref{fig:kagome-curvature}(b).

The Berry connection associated with these three gauge choices are also shown in Fig.~\ref{fig:kagome-connection}. Similar to the case in the checker board lattice model, the position of the vortex strongly changes the distribution of $\mathbf{A}(\vk)$ field.

\subsection{Twisted bilayer transition metal dichalcogenides}

\begin{figure}[t]
    \centering
    \includegraphics[width=0.75\linewidth]{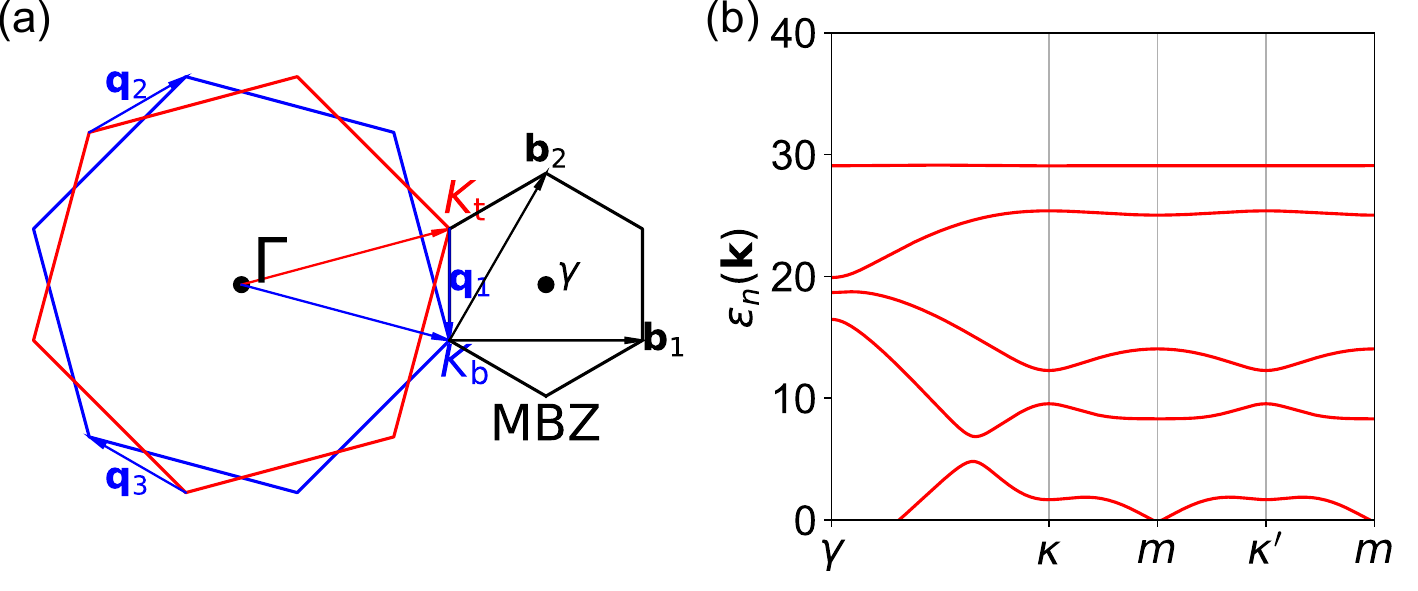}
    \caption{(a) The moir\'e Brillouin zone. Red and blue hexagons stand for the single layer Brillouin zones of top and bottom layers, respectively. The small black hexagon is the moir\'e Brillouin zone. (b) The band structure of the AA-stacked twisted bilayer $\rm WSe_2$ at twisting angle $\theta = 1.43^\circ$. The top-most energy band carries Chern number $\mathcal{C} = -1$. Dispersion energy is measured in meV. The model parameters are the same as in Ref.~\cite{Devakul2021Magic}.}
    \label{fig:tmd-bands}
\end{figure}

\begin{figure}[t]
    \centering
    \includegraphics[width=0.5\linewidth]{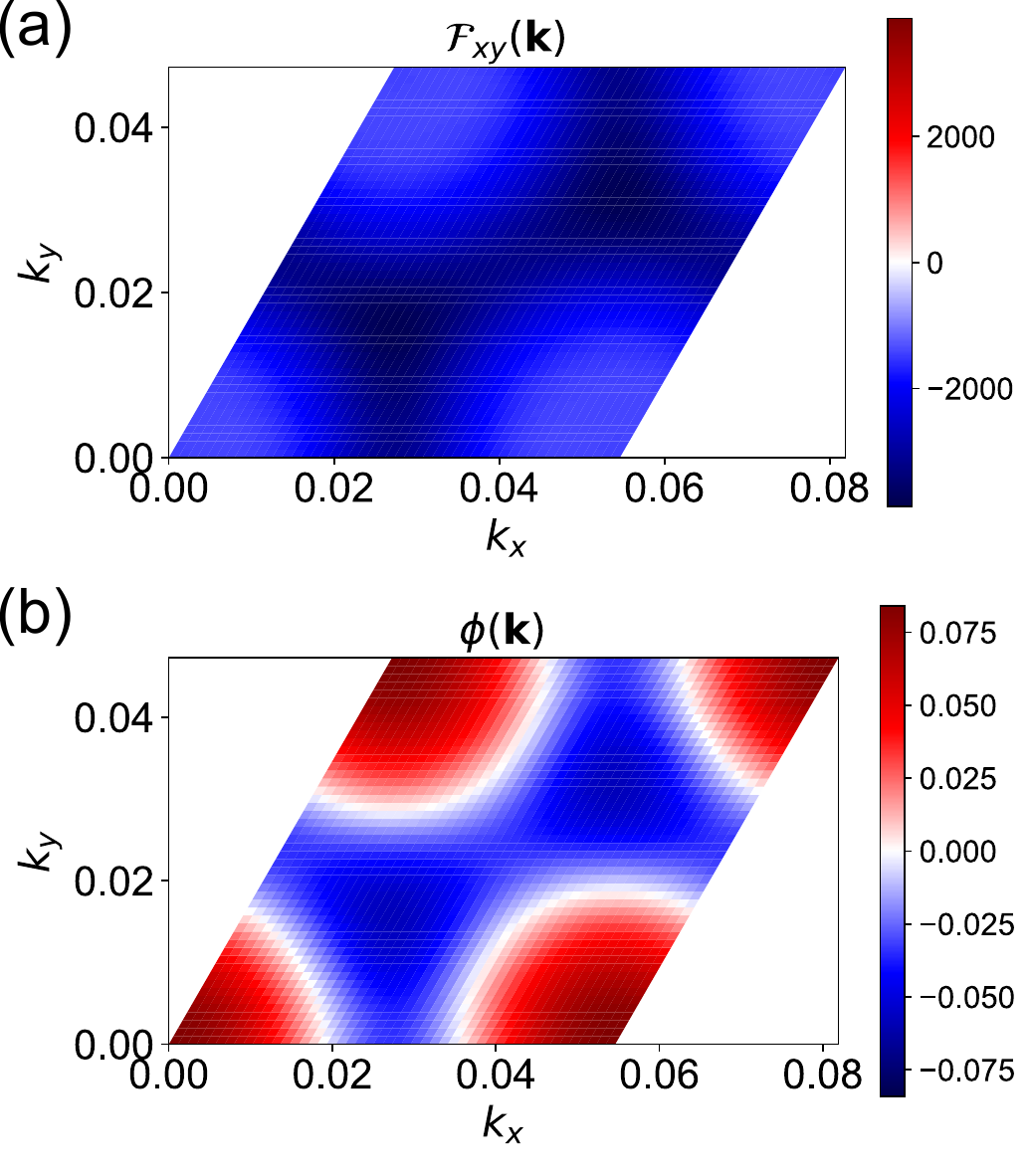}
    \caption{(a) The Berry curvature distribution of the top most moir\'e band in twisted bilayer $\rm WSe_2$.
    (b) The smooth vortex potential of this Chern band.
    In both subfigures, the momentum components $k_x$, $k_y$ are measured using \AA$^{-1}$.
    }
    \label{fig:tmd-curvature}
\end{figure}

\begin{figure}[t]
    \centering
    \includegraphics[width=0.75\linewidth]{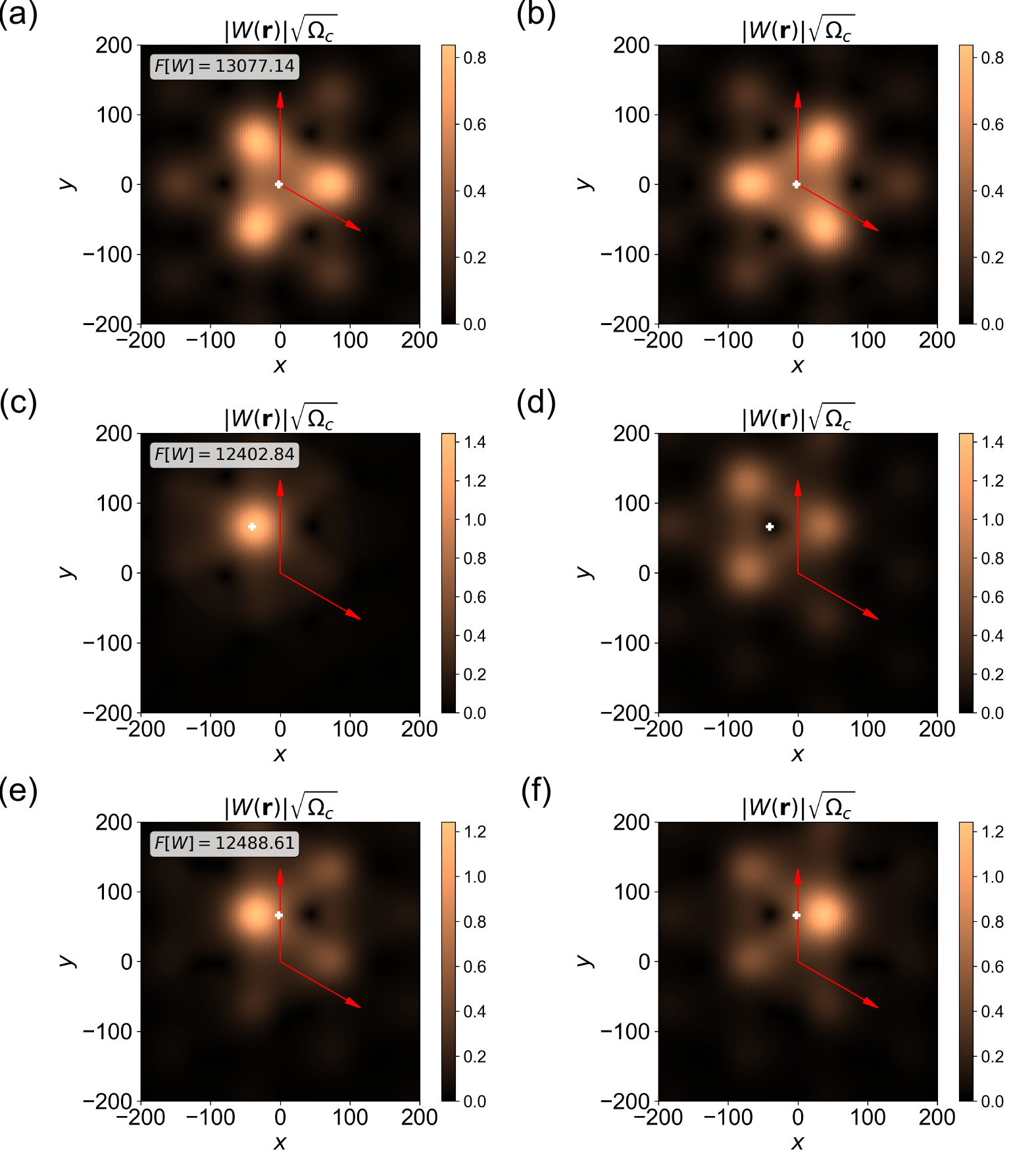}
    \caption{(a-b) Wannier functions in top layer (a) and bottom layer (b) of the top-most moir\'e Chern band in twisted bilayer $\rm WSe_2$ with vortex at $\gamma$ point $\vk_v = \mathbf{0}$. 
    In (c-d) and (e-f), the vortices are at $\kappa$ and $m$ points with $\vk_v = (\mathbf{b}_1 + \mathbf{b}_2)/3$ and $\vk_v = \mathbf{b}_1/2$, respectively. 
    In each subfigure, the red arrows stand for the Bravais basis vectors, and the real space coordinates $x$ and $y$ are measured by \AA.
    The values of $F[W]$, measured in \AA$^2$, are evaluated over $24\times 24$ moir\'e unit cells with $30\times 30$ points in each unit cell.
    }
    \label{fig:tmd-wannier}
\end{figure}

\begin{figure}[t]
    \centering
    \includegraphics[width=0.5\linewidth]{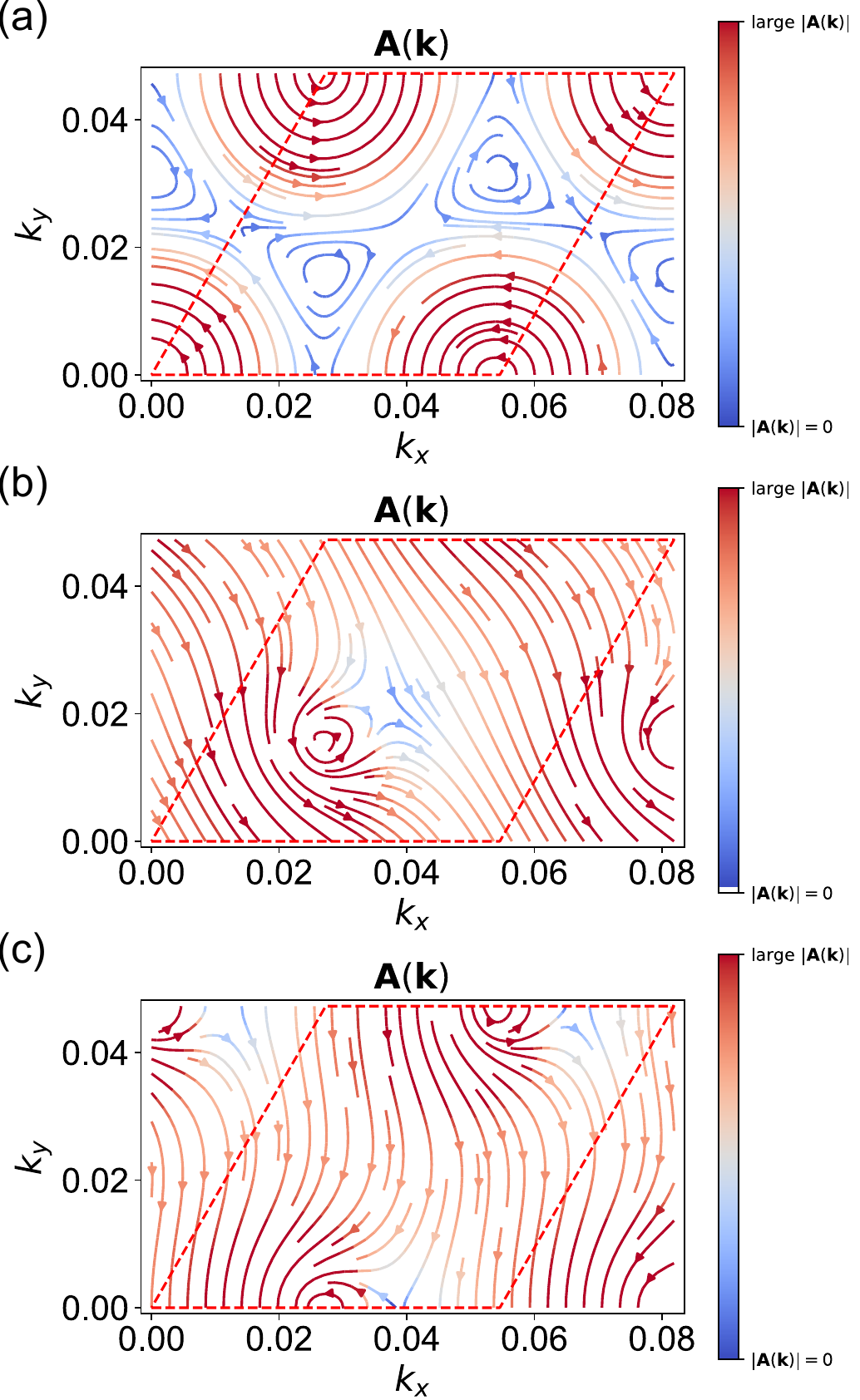}
    \caption{The Berry connection of the top-most moir\'e Chern band in twisted bilayer $\rm WSe_2$ with vortex located at (a) $\gamma$, (b), $\kappa$ and (c) $m$ points. }
    \label{fig:tmd-connection}
\end{figure}

The gauge fixing procedure is not limited to the Bloch states in tight-binding models. In this subsection, we provide an example for continuum model with a periodic potential, namely the twisted bilayer transition metal dichalcogenides (TMD), which can host nearly flat Chern bands with properly tuned twisting angles \cite{Wu2019topological,Devakul2021Magic}.

The single layer Brillouin zone of TMD in its $2H$ structural phase is hexagonal, as shown in Fig.~\ref{fig:tmd-bands}(a).
Around the $K$ and $K'$ points, the band structure forms spin-polarized hole pockets due to the strong spin-orbital coupling. Hence, only one spin species is active near the $K$ (or $K'$) point, which is usually refereed as {\it spin-valley locking} \cite{Liu2013Three,Kormanyos2015kp}. 
Therefore, its low-energy physics could be well-described using a quadratic band with an effective mass $m^*$.

When the bilayer system is twisted by a small angle $\theta$, the $K$ and $K'$ points from the two layers will be separated from each other by $q_\theta = 8\pi\sin(\theta/2)/3a_0$, in which $a_0$ is the single layer lattice constant.
The value of $q_\theta$ controls the size of the moir\'e Brillouin zone. 
As depicted in Fig.~\ref{fig:tmd-bands}(a), the reciprocal basis vectors of the moir\'e lattice are given by $\mathbf{b}_1 = q_\theta (\sqrt{3},0)$ and $\mathbf{b}_2 = q_\theta(\sqrt{3}/2,3/2)$. 

With the periodic moir\'e potential considered, the single valley effective Hamiltonian will have the following form:
\begin{equation}\label{eqn:tmd-hamiltonian}
    H = \left(
        \begin{array}{cc}
            \frac{\nabla^2}{2m^*} + V_+(\mathbf{r}) & T(\mathbf{r})\\
            T^*(\mathbf{r}) & \frac{\nabla^2}{2m^*} + V_-(\mathbf{r})
        \end{array}
    \right)\,.
\end{equation}
in which $V_\pm(\mathbf{r})$ and $T(\mathbf{r})$ stand for intra-layer moir\'e potential and inter-layer tunneling.
This Hamiltonian is constrained by the single valley symmetry group of TMD generated by $C_{3z}$ and $C_{2y}T$ operations.
Using the lowest-harmonic approximation \cite{Wu2019topological}, the intra-layer potentials will have the following form:
\begin{equation}
    V_{\ell}(\mathbf{r}) = 2V \sum_{j=1,3,5} \cos(\mathbf{g}_j\cdot\mathbf{r} + \ell \psi)\,,
\end{equation}
in which the reciprocal vectors are defined as $\mathbf{g}_1 = -\mathbf{g}_4 = \mathbf{b}_1$, $\mathbf{g}_3 = -\mathbf{g}_6 = \mathbf{b}_2 - \mathbf{b}_1$ and $\mathbf{g}_5 = -\mathbf{g}_2 = -\mathbf{b}_2$.
Due to the $C_{2y}T$ symmetry, the top $(\ell=+)$ and bottom layer $(\ell=-)$ potentials are differed by the sign of phase factor $\psi$.
The lowest harmonic form of the inter layer tunneling can be written as:
\begin{equation}
    T(\mathbf{r}) = w\sum_{j=1}^3 e^{i\vq_j\cdot \mathbf{r}}\,,
\end{equation}
in which $w \in \mathbb{R}$ due to $C_{3z}$ symmetry, and the three $\vq_j$ vectors are defined as labeled in Fig.~\ref{fig:tmd-bands}(a).
Fourier transforming $V_\pm(\mathbf{r})$ and $T(\mathbf{r})$ into reciprocal space, one can easily rewrite Eq.~(\ref{eqn:tmd-hamiltonian}) into the form of Eq.~(\ref{eqn:def-pw-hamiltonian}), and compute its band structure and Bloch wave functions.

In $\rm WSe_2$ bilayer, the effective band mass is around $m^* \approx 0.43m_e$, the intra-layer potential is around $V \approx 9\,\rm meV$, $\psi \approx 128^\circ$, and the inter-layer tunneling strength is around $w \approx 18\,\rm meV$ \cite{Devakul2021Magic}.
Assuming the twisting angle is $\theta = 1.43^\circ$, we computed the band structure of twisted bilayer $\rm WSe_2$, which is shown in Fig.~\ref{fig:tmd-bands}(b). A nearly flat band can be found at the top of its band structure. 
The Berry curvature evaluated on a $48\times 48$ momentum mesh, shown in Fig.~\ref{fig:tmd-curvature}(a), indicates that it carries Chern number $\mathcal{C} = -1$. Solving the Poisson's equation yields the smooth vortex potential, which is presented in Fig.~\ref{fig:tmd-curvature}(b). The value $\phi(\vk)$ approaches its maximum value at $\gamma$ point, and its minimum value at $\kappa$ and $\kappa'$ points.

We then fix the gauge of the Bloch states following the method described in Sec.~\ref{sec:gauge} and use Eq.~(\ref{eqn:wannier-pw}) to compute the Wannier functions, which have two components in top and bottom layers. When the vortex is placed at $\gamma$ point, the Wannier functions in both layers, computed on a $24\times 24$ lattice, are presented in Figs.~\ref{fig:tmd-wannier}(a-b). Wave functions form clover shape orbitals, and the charge density in the two layers concentrate in $MX$ and $XM$ stacking regions, respectively \cite{Devakul2021Magic}. By shifting the vortex to the $\kappa$ point, the center of the Wannier function will be moved to the center of $MX$ stacking region, which is provided in Figs.~\ref{fig:tmd-wannier}(c-d). 
Moreover, we observe a significant asymmetry in the wave function between the two layers, differing from the scenario observed in the presence of the vortex at $\gamma$. The charge density concentrates primarily on a single $MX$ stacking position in the top layer, while retaining the clover shape in the bottom layer.
Finally, we show the Wannier function with vortex moved to the $m$ point in Figs.~\ref{fig:tmd-wannier}(e-f). The charge center becomes the middle point between two neighboring $MX$ and $XM$ stacking positions, albeit the charge density in both layers are still mostly concentrated within these $MX$ and $XM$ stacking regions.

Since the computation cost of $F[W]$ in continuum models are usually much larger than tight-binding models, we only computed $F[W]$ with $\vk_v$ at $\gamma$, $\kappa$ and $m$ points. 
Since the Chern number is $\mathcal{C} = -1 < 0$, we expect that the minimum value of $F[W]$ is reached when $\vk_v$ is at the minimum of $\phi(\vk)$, which is opposite to the previous examples.
The values of $F[W]$ are labeled in Figs.~\ref{fig:tmd-wannier}(a,c,e). Among these three cases, the Wannier function with $\vk_v$ at $\gamma$ has the largest $F[W]$, while $\vk_v$ at $\kappa$ has the smallest $F[W]$, consistent with the prediction of $\phi(\vk)$. 

Lastly, we also provide the distribution of the Berry connection in Fig.~\ref{fig:tmd-connection} for reference. Qualitatively, the $\mathbf{A}(\vk)$ field in this moir\'e Chern band is similar to the Chern band in the kagome lattice model.

\subsection{Ideal Chern bands}\label{sec:ideal-chern-band}

\begin{figure}[t]
    \centering
    \includegraphics[width=0.5\linewidth]{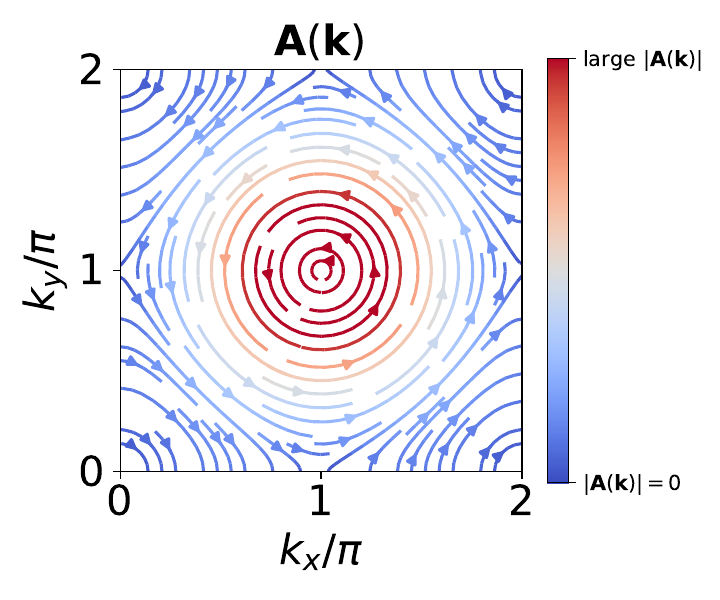}
    \caption{Berry connection of the LLL derived from Eq.~(\ref{eqn:ideal-connection}). Here the magnetic unit cell basis vectors are chosen as $\mathbf{a}_1 = (1, 0)$, $\mathbf{a}_2 = (0, 1)$, and magnetic field $B = 2\pi$. The $\mathbf{A}(\vk)$ field diverges at $\vk_v = (\pi, \pi)$.}
    \label{fig:lll-connection}
\end{figure}

A Chern band is considered to be an \emph{ideal Chern band}, if the inequality in Eq.~(\ref{eqn:trace-bound}) is saturated:
\begin{equation}\label{eqn:trace-condition}
    {\rm Tr}\,g(\vk) = |\mathcal{F}_{xy}(\vk)|\,.
\end{equation}
This condition is also known as the \emph{trace condition}.
It has been shown in Ref.~\cite{claassen_position-momentum_2015} that if the (unnormalized) Bloch states $u_{\alpha, n}(\vk)$ can be written as meromorphic/anti-meromorphic functions of $k = k_x + i k_y$, the trace condition is guaranteed to be satisfied. 

Ideal Chern bands can also be interpreted as generalization of Landau levels in some context \cite{Estienne-ideal-2023}.
Flat moir\'e bands in chiral twisted bilayer graphene and twisted bilayer TMDs have been shown to be closely related with ideal Chern bands \cite{tarnopolsky_origin_2019,ledwith_fractional_2020, dong_composite_2023,Devakul2021Magic,Wu2019topological}. Moreover, FCI states, including Laughlin states are exact ground states for interacting electrons moving in these ideal flat Chern bands with short-range interactions \cite{ledwith_fractional_2020,wang_exact_2021,wang2023origin}, due to the exact Girvin-MacDonald-Platzman algebra in these systems \cite{girvin1985collective,wang2023origin}. 

We will show that the Bloch states of an ideal Chern band, written in its holomorphic form, already satisfies the Coulomb gauge condition, thereby the corresponding Wannier function being well localized with power-law tails. 
The holomorphic condition can be represented by the Cauchy-Riemann equation:
\begin{equation}\label{eqn:holomorphic-bloch}
    \frac{\partial}{\partial k^*}\tilde{u}_\vk(\mathbf{r}) = 0\,,
\end{equation}
in which $k^* = k_x - ik_y$ is the (conjugate) complex number representation of Bloch momentum, and $\tilde{u}_\vk(\mathbf{r})$ is the unnormalized Bloch function:
\begin{equation}
    u_\vk(\mathbf{r}) = \frac{1}{\sqrt{\mathcal{N}(\vk)}}\tilde{u}_\vk(\mathbf{r})\,.
\end{equation} 
Here $\mathcal{N}(\vk) \in \mathbb{R}^+ \cup \{0\}$ is the normalization factor, which is a function of $\vk$ but not $\mathbf{r}$. 
Obviously, the holomorphic condition of the Bloch states can also be interpreted as a special gauge choice.
Under such gauge choice,  the Berry connection can be written as the following form:
\begin{align}
    \mathbf{A}(\vk) =& -i \int_{\Omega_c}d^2r\, u^*_\vk(\mathbf{r})\nabla_{\vk}u_\vk(\mathbf{r})\nonumber\\
    =& -i\frac{1}{\mathcal{N}(\vk)}\int_{\Omega_c}d^2r\,\tilde{u}^*_{\vk}(\mathbf{r})\nabla_\vk \tilde{u}_\vk(\mathbf{r}) + \frac{i}{2}\nabla_\vk \log \mathcal{N}(\vk)\,.
\end{align}
Its holomorphic component $A_x + iA_y$ can be simplified into:
\begin{align}
    A_x + i A_y =&  -i\frac{2}{\mathcal{N}(\vk)}\int_{\Omega_c}d^2r\,\tilde{u}^*_{\vk}(\mathbf{r})\left(\frac{\partial}{\partial k^*}\right)\tilde{u}_\vk(\mathbf{r}) + i\frac{\partial}{\partial k^*} \log \mathcal{N}(\vk)\nonumber\\
    =& i\frac{\partial}{\partial k^*} \log \mathcal{N}(\vk)\,.\label{eqn:holomorphic-connection}
\end{align}
Here the first term vanishes due to Eq.~(\ref{eqn:holomorphic-bloch}). Since the Berry connection is defined as real-valued vectors, the two components $A_x$ and $A_y$ are given by the real and imaginary parts of Eq.~(\ref{eqn:holomorphic-connection}), respectively:
\begin{align}
    A_x &= {\rm Re}\left(i\frac{\partial}{\partial k^*} \log \mathcal{N}(\vk)\right) = -\frac{1}{2}\frac{\partial}{\partial k_y} \log\mathcal{N}(\vk)\,,\\
    A_y &= {\rm Im}\left(i\frac{\partial}{\partial k^*} \log \mathcal{N}(\vk)\right) = \frac{1}{2}\frac{\partial}{\partial k_x}\log\mathcal{N}(\vk)\,.
\end{align}
These two equations can indeed be rewritten into a compact form:
\begin{equation}\label{eqn:ideal-connection}
    \mathbf{A}(\vk) = \frac{\hat{\mathbf{z}}}{2}\times \nabla_\vk \log \mathcal{N}(\vk)\,,
\end{equation}
and its corresponding Berry curvature is simply given by:
\begin{equation}
    \mathcal{F}_{xy}(\vk) = \frac{1}{2}\nabla_\vk^2 \log \mathcal{N}(\vk)\,.
\end{equation}
It is also easy to check that the Berry connection of such gauge choice automatically satisfies the Coulomb gauge condition:
\begin{equation}
    \nabla_\vk \cdot \mathbf{A}(\vk) = \left[-\frac{1}{2}\frac{\partial}{\partial k_x}\frac{\partial}{\partial k_y} + \frac{1}{2}\frac{\partial }{\partial k_y}\frac{\partial}{\partial k_x}\right]\log\mathcal{N}(\vk) = 0\,.
\end{equation}
Obviously, zeros and singularities of the normalization function $\mathcal{N}(\vk)$ generate vortices of the Berry connection.

We conclude that the gauge choices of the Bloch wave functions in an ideal Chern band does not need to be further fixed, if its holomorphic form is already known. By simply normalizing the holomorphic Bloch wave function, one can readily determine its optimal gauge choice, which corresponds to minimizing $F[W]$.

As an example, we provide the expression of the normalization factor of the lowest Landau level on a 2D torus, which is one of the simplest ideal Chern bands \cite{Rashba1997Orthogonal,Panfilov2016chiral}:
\begin{equation}\label{eqn:lll-norm}
    \mathcal{N}(\mathbf{k}) = \sqrt{\frac{2 a_1}{a_2}}e^{-\frac{a_1a_2 k_y^2}{2\pi}}\vartheta_3\left(\frac{k}{2} a_1\Big{|}i\frac{a_1}{a_2}\right)\vartheta_3\left(\frac{k^*}{2} a_1\Big{|}i\frac{a_1}{a_2}\right)\,,
\end{equation}
in which $\vartheta_3$ stands for the Jacobi theta function, and $a_1, a_2$ are the lengths of the rectangular magnetic unit cell satisfying the condition $Ba_1a_2 = 2\pi$.
The corresponding Berry connection field can be easily solved numerically as shown in Fig.~\ref{fig:lll-connection}. 
Due to the Perelomov overcompleteness condition \cite{Perelomov1971On}, the normalization factor vanishes at $\vk_v = (\pi, \pi)$, which leads to a vortex of $\mathbf{A}(\vk)$ at this position.

\section{Projected interacting Hamiltonian}\label{sec:interaction}

With these Wannier functions, we are able to study the interacting Hamiltonian projected in these flat bands using a real space basis. Traditionally, real space basis for projected interacting Hamiltonian are obtained in trivial systems. 
Wannier representation for lowest Landau levels defined on a torus \cite{Rashba1997Orthogonal} leads to discussion bosonic topological ordered states~\cite{Burkov2010fractional,Panfilov2016chiral}.
And recently the interacting matrix elements under the Wannier bases in twisted bilayer graphene has also been studied \cite{zang_real_2022}.
However, it is more common to project the interaction terms into topological bands using the Bloch basis, and the real space aspect of interacting electrons in topological bands are still relatively under-explored. 

\subsection{Interacting spinless fermion}

In this section, we study the projected interactions in the case of spinless fermions in a tight-biding setting. A generic density-density interaction in a tight-binding model will have the following form:
\begin{equation}
    H_{\rm int} = \frac12 \sum_{\mathbf{R}\alpha,\mathbf{R'}\beta}\mathcal{U}_{\alpha\beta}(\mathbf{R})n_{\mathbf{R} + \mathbf{R}',\alpha}n_{\mathbf{R}',\beta}\,,
\end{equation}
where $\mathcal{U}_{\alpha\beta}(\mathbf{R}) \in \mathbb{R}$ is the interaction potential between two electrons located at $\mathbf{R} + \bm{\tau}_\alpha$ and $\bm{\tau}_\beta$, and $n_{\mathbf{R},\alpha,s} = c^\dagger_{\mathbf{R},\alpha,s}c_{\mathbf{R},\alpha,s}$ is the local particle number operator. By definition, the interaction potential satisfies $\mathcal{U}_{\alpha\beta}(\mathbf{R}) = \mathcal{U}_{\beta\alpha}(-\mathbf{R})$. 
This Hamiltonian can also be written in Bloch electron operators:
\begin{align}
    &H_{\rm int} = \sum_{\substack{\vk\vk'\vq\\ \alpha\beta}}\frac{\tilde{\mathcal{U}}_{\alpha\beta}(\vq)}{2N_1N_2}c^\dagger_{\vk + \vq,\alpha}c_{\vk,\alpha}c^\dagger_{\vk'-\vq,\beta}c_{\vk',\beta}\,,\\
    & \tilde{\mathcal{U}}_{\alpha\beta}(\vq) = \sum_{\mathbf{R}}\mathcal{U}_{\alpha\beta}(\mathbf{R})e^{-i\vq\cdot(\mathbf{R} + \bm{\tau}_\alpha - \bm{\tau}_\beta)}
\end{align}
Similar to the definition of Bloch Hamiltonian $h(\vk)$, the Fourier transformed interaction potential $\tilde{\mathcal{U}}_{\alpha\beta}(\vq)$ is not periodic in the reciprocal space:
\begin{equation}
    \tilde{\mathcal{U}}_{\alpha\beta}(\vq + \mathbf{G}) = \tilde{\mathcal{U}}_{\alpha\beta}(\vq)e^{-i\mathbf{G}\cdot(\bm{\tau}_\alpha - \bm{\tau}_\beta)}\,.
\end{equation}

Our focus here is the strongly correlated effects in an active topological band. Thus, it is more convenient to use Eq.~(\ref{eqn:band-operator}) and rewrite the interacting Hamiltonian using the Bloch band creation/annihilation operators $\gamma^\dagger_{\vk,n}, \gamma_{\vk,n}$. 
We then project the many-body Hamiltonian into the $n$-th band, and consequently the projected interacting Hamiltonian can be written as the following form: 
\begin{align}
    & \overline{H}_{\rm int} = \frac{1}{2N_1N_2}\sum_{\substack{\vk\vk'\vq}}\tilde{\mathcal{V}}(\vq;\vk,\vk')\gamma^\dagger_{\vk + \vq, n}\gamma^\dagger_{\vk' - \vq, n}\gamma_{\vk', n}\gamma_{\vk, n}\,,\label{eqn:spinless-proj-bloch}\\
    & \tilde{\mathcal{V}}(\vq;\vk,\vk') = \sum_{\alpha\beta}\tilde{\mathcal{U}}_{\alpha\beta}(\vq)M_{\alpha}(\vq, \vk)M_\beta(-\vq, \vk') \,,\label{eqn:bloch-proj-interaction}\\
    & M_\alpha(\vq, \vk) = u^*_{\alpha,n}(\vk + \vq) u_{\alpha,n}(\vk)\,.
\end{align}
It can be shown that the projected interaction matrix elements $\tilde{\mathcal{V}}(\vq;\vk,\vk')$ is periodic in all three momentum variables. 

We also note that there are $\sim(N_1N_2)^3$ independent components in $\tilde{\mathcal{V}}$ due to the form factors $M_\alpha(\vq, \vk)$ in the projection procedure. In contrast, the unprojected interaction matrix elements $\mathcal{U}_{\alpha\beta}(\vq)$ only have $\sim N_1N_2$ components. 

\begin{table}[]
    \centering
    \begin{tabular}{c|c|c|c r l}
    \toprule
    $\overline{H}_{int}$ &\textbf{Channel} & \textbf{\# Centers} &${\cal V}({\mathbf R};{\mathbf d},{\mathbf d}')$ &\textbf{Second quantized form} & \\
    \hline
    $\overline{H}_V$ &Direct &2 &${\cal V}({\mathbf R};{\mathbf 0},{\mathbf 0})$ &$w_{\mathbf{R}+\mathbf{R}_0}^\dagger w_{\mathbf{R}_0}^\dagger w_{\mathbf{R}_0} w_{\mathbf{R}+\mathbf{R}}$ &$= n_{\mathbf{R}_0} n_{\mathbf{R}+\mathbf{R}_0}$ \\
    \hline
    $\overline{H}_X$ &Exchange &2  &${\cal V}({\mathbf R};-{\mathbf R},{\mathbf R})$ &$w_{\mathbf{R}_0}^\dagger w_{\mathbf{R}+\mathbf{R}_0}^\dagger w_{\mathbf{R}_0} w_{\mathbf{R}+\mathbf{R}_0} $ &$= -n_{\mathbf{R}_0} n_{\mathbf{R}+\mathbf{R}_0}$ \\
    \hline
    $\overline{H}_{A_h}$ &Assisted hopping &3  &${\cal V}({\mathbf R};{\mathbf d},{\mathbf 0})$ &$w^\dagger_{\mathbf{R} + \mathbf{d} +\mathbf{R}_0}w^\dagger_{\mathbf{R}_0} w_{\mathbf{R}_0}w_{\mathbf{R} + \mathbf{R}_0} $ &$= -w^\dagger_{\mathbf{R} + \mathbf{d} +\mathbf{R}_0}w_{\mathbf{R} + \mathbf{R}_0}n_{\mathbf{R}_0}$ \\
    & &  &${\cal V}({\mathbf R};{\mathbf 0},{\mathbf d})$ &$w^\dagger_{\mathbf{R}+\mathbf{R}_0}w^\dagger_{\mathbf{d} + \mathbf{R}_0} w_{ \mathbf{R}_0}w_{\mathbf{R} + \mathbf{R}_0} $ &$= w^\dagger_{\mathbf{d} + \mathbf{R}_0} w_{\mathbf{R}_0}n_{\mathbf{R} + \mathbf{R}_0}$ \\
    \hline
    $\overline{H}_{A_{ex}}$ &Assited exchange &3  &${\cal V}({\mathbf R};-{\mathbf R},{\mathbf d})$ &$w^\dagger_{\mathbf{R}_0}w^\dagger_{ \mathbf{d} +\mathbf{R}_0} w_{\mathbf{R}_0}w_{\mathbf{R} + \mathbf{R}_0} $ &$= -w^\dagger_{ \mathbf{d} +\mathbf{R}_0}w_{\mathbf{R} + \mathbf{R}_0}n_{\mathbf{R}_0}$ \\
    & & &${\cal V}(-{\mathbf R};{\mathbf d}+{\mathbf R},-{\mathbf R})$ &$w^\dagger_{\mathbf{d} +\mathbf{R}_0}w^\dagger_{ -\mathbf{R}+\mathbf{R}_0} w_{\mathbf{R}_0}w_{-\mathbf{R} + \mathbf{R}_0} $ &$= -w^\dagger_{\mathbf{d} +\mathbf{R}_0} w_{\mathbf{R}_0}n_{-\mathbf{R} + \mathbf{R}_0}$ \\
    \hline
    $\overline{H}_R$ &Ring exchange &4  &${\cal V}({\mathbf a};-{\mathbf a}',{\mathbf a}')\,,\ldots$  &$w^\dagger_{\mathbf{a}-\mathbf{a}'+\mathbf{R}_0}w^\dagger_{\mathbf{a}' + \mathbf{R}_0} w_{\mathbf{R}_0}w_{\mathbf{a}+\mathbf{R}_0}$ & \\
    \hline
    $\overline{H}_{\rm extra}$  &Extra terms &4  &${\cal V}({\mathbf R};{\mathbf d},{\mathbf d}')$  &$w^\dagger_{\mathbf{R} + \mathbf{d} +\mathbf{R}_0}w^\dagger_{\mathbf{d}' + \mathbf{R}_0} w_{\mathbf{R}_0}w_{\mathbf{R} + \mathbf{R}_0}$ & \\
    \toprule
    \end{tabular}
    \caption{
    Channels of the projected interacting Hamiltonians in the Wannier basis in second qunatized form Eq.~(\ref{eqn:spinless-proj-wannier}). Here ${\mathbf R}_0$ is the reference point; ${\mathbf R}$, ${\mathbf d}$ and ${\mathbf d}'$ are not equal if not specified; ${\mathbf a}$, ${\mathbf a}'$ and ${\mathbf a}-{\mathbf a}'$ are vectors pointing from ${\mathbf R}_0$ to its nearest sites. $n_{\mathbf{x}}=w_{\mathbf{x}}^\dagger w_{\mathbf{x}}$ is the charge density operator for the Wannier basis. The sketches of these channels are shown in 
    Fig.~3(a) of the main text and in Fig.~\ref{fig:spinless-interaction-channels}. }
    \label{tab:projected_H_channels}
\end{table}

Equivalently, such projected interacting Hamiltonian can also be expressed using the Wannier states:
\begin{align}
    \overline{H}_{\rm int} =& \frac{1}{2}\sum_{\mathbf{R}_0,\mathbf{R}\mathbf{d}\mathbf{d}'}\mathcal{V}(\mathbf{R};\mathbf{d}, \mathbf{d}') w^\dagger_{\mathbf{R} + \mathbf{d} +\mathbf{R}_0, n}w^\dagger_{\mathbf{d}' + \mathbf{R}_0, n} w_{\mathbf{R}_0,n}w_{\mathbf{R} + \mathbf{R}_0,n}\,,\label{eqn:spinless-proj-wannier}
\end{align}
in which the projected interaction matrix elements are given by the overlap integral of the Chern band Wannier functions:
\begin{align}
    &\mathcal{V}(\mathbf{R};\mathbf{d},\mathbf{d}') \nonumber\\
    =& \frac{1}{(N_1N_2)^3}\sum_{\vk\vk'\vq}e^{i\vq\cdot(\mathbf{R} + \mathbf{d} - \mathbf{d}')}e^{i\vk\cdot \mathbf{d}}e^{i\vk'\cdot \mathbf{d}'} \tilde{\mathcal{V}}(\vq;\vk,\vk') \\
    =& \sum_{\mathbf{R}_1\mathbf{R}_2\alpha\beta}\mathcal{U}_{\alpha\beta}(\mathbf{R}_1 - \mathbf{R}_2) W^{*}_{\alpha}(\mathbf{R}_1 - \mathbf{R} - \mathbf{d})W_\alpha(\mathbf{R}_1 - \mathbf{R}) W^{*}_\beta(\mathbf{R}_2-\mathbf{d}') W_\beta(\mathbf{R}_2)\,.
\end{align}
Similar to the projected Hamiltonian expressed using Bloch states in Eq.~(\ref{eqn:spinless-proj-bloch}), there are $\sim(N_1N_2)^3$ independent components because of the non-trivial form factors of the Wannier functions. Hence, more complicated structures, for example, multi-center interaction including ring-exchange terms, can emerge in projected interacting Hamiltonians.

These terms can be classified into different categories:
\begin{itemize}
    \item One-center term: $\mathbf{R} = \mathbf{d} = \mathbf{d}' = \mathbf{0}$; in spinless one-band systems, these terms are equivalent to a chemical potential shift, which only has trivial physical effects. Hence, we do not take these terms into consideration.
    \item Two-center terms can be classified into three types: direct channel, exchange channel and pair hopping channel.

    The direct channel Hamiltonian contains the terms with $\mathbf{d}=\mathbf{d}'=\mathbf{0}$:
        \begin{equation}
            V(\mathbf{R}) = \mathcal{V}(\mathbf{R};\mathbf{0},\mathbf{0})\,,~~\mathbf{R} \neq \mathbf{0}\,.
        \end{equation}
    The exchange channel contains terms that swaps the second and the fourth fermion operators, which can be written as:
        \begin{equation}
            X(\mathbf{R}) = \mathcal{V}(\mathbf{R}; -\mathbf{R}, \mathbf{R})\,,~~\mathbf{R} \neq \mathbf{0}\,.
        \end{equation}
    Pair hopping contains interaction terms that have two creation operators at the same position, and two annihilation operators at the same position, which can be written as:
        \begin{equation}
            P(\mathbf{R}) = \mathcal{V}(\mathbf{0};\mathbf{R},\mathbf{R})\,,~~\mathbf{R}\neq \mathbf{0}\,.
        \end{equation}
    However, these terms are irrelevant to spinless fermion systems due to Pauli's exclusion principle.

    In the flat band limit, these two-center terms, although being long-range interactions, usually commute with the particle number operators associated with Wannier states. Therefore, they can be diagonalized in the Fock basis of Wannier states, and cannot include any ``quantum'' effects.

    \item Tri-center terms are also possible in the projected interacting Hamiltonian. These terms can be roughly classified into two categories: assisted hopping $A_{h}$, assisted exchange $A_{ex}$ and assisted pairing $A_p$ terms.
    
    \begin{align}
        A_h^1(\mathbf{R}, \mathbf{d}) &= \mathcal{V}(\mathbf{R};\mathbf{d},\mathbf{0})\,,\\
        A_h^2(\mathbf{R}, \mathbf{d}) &= \mathcal{V}(\mathbf{R};\mathbf{0},\mathbf{d})\,.
    \end{align}

    Similar to the exchange and pair hopping interactions in two-center terms, there are also assisted ``exchange'' interactions:
    \begin{align}
        A^1_{ex}(\mathbf{R}, \mathbf{d}) &= \mathcal{V}(\mathbf{R};-\mathbf{R},\mathbf{d})\,,\\
        A^2_{ex}(\mathbf{R}, \mathbf{d}) &= \mathcal{V}(-\mathbf{R};\mathbf{d}+\mathbf{R},-\mathbf{R})\,,
    \end{align}
    and assisted ``pairing'' interactions:
    \begin{align}
        A_p^1(\mathbf{R}, \mathbf{d}) &= \mathcal{V}(\mathbf{d};\mathbf{R}-\mathbf{d},\mathbf{R})\,,\\
        A_p^2(\mathbf{R}, \mathbf{d}) &= \mathcal{V}(\mathbf{0};\mathbf{R} + \mathbf{d}, \mathbf{R})\,.
    \end{align}
    Similar to the pair hopping terms $X$, $A_p$ terms will have no physical effect in spinless fermion systems due to Pauli's exclusion principle.

    \item Four-center interactions are generic terms in which fermionic operators are located at four different positions. We can simply separate them into two categories based on distance. When the four operators are sitting on the four corners of a Bravais lattice plaquette, we refer to these terms as ``ring-exchange'' \cite{Burkov2010fractional}, and the corresponding Hamiltonian as $\overline{H}_R$.

    Other four-center terms with longer operator separation distances will be denote as $\overline{H}_{\rm extra}$.
\end{itemize}
In summary, a projected interacting Hamiltonian in a spinless flat band contains the following six types of terms, and can be written in the following form:
\begin{equation}
    \overline{H}_{\rm int} = \overline{H}_V + \overline{H}_X + \overline{H}_{A_h} + \overline{H}_{A_{ex}} + \overline{H}_R + \overline{H}_{\rm extra}\,.
\end{equation}
Fig.~\ref{fig:spinless-interaction-channels} provides sketches for different interaction channels.

Conventionally, only two-center terms are considered in the projected interacting models, for example, in Hubbard-Kanamori Hamiltonian \cite{Kanamori1963,Georges2013Strong}. In trivial bands, this assumption is usually justified, since its Wannier function is exponentially localized.
Even so, the effect of assisted hoppings terms in trivial bands has recently been demonstrated to be significant \cite{Jiang2023Density}. 
In topological bands, the Wannier functions are power-law localized, and the long-range interactions can be more important. 

\begin{figure}[t]
    \centering
    \includegraphics[width=0.75\linewidth]{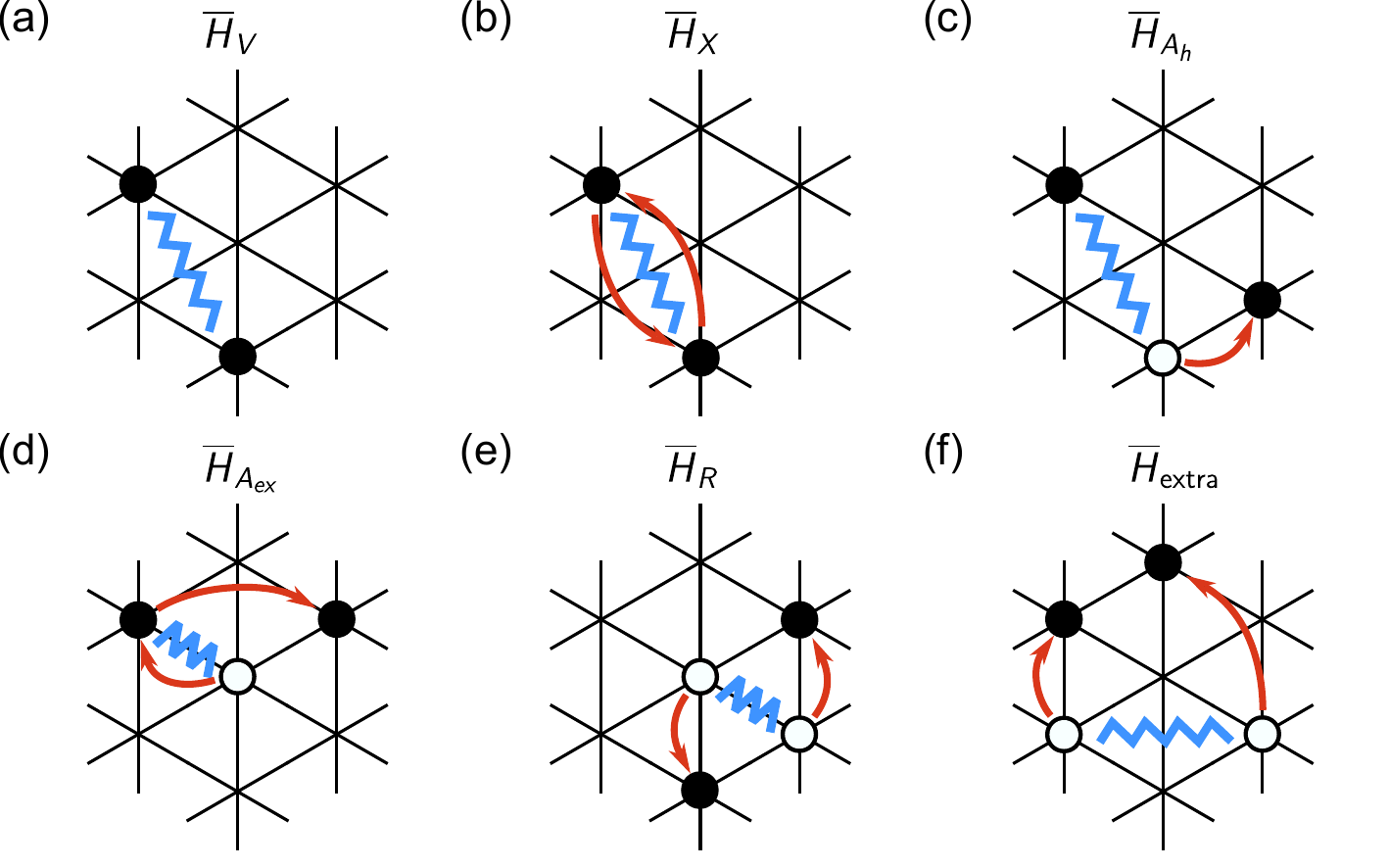}
    \caption{Sketches for different channels in projected interacting Hamiltonian of a spinless fermion model. Filled and empty circles stand for the location of creation and annihilation operators, respectively. The blue wavy lines and the red arrows represent the vectors $\mathbf{R}$, $\mathbf{d}$ and $\mathbf{d}'$. The first two channels (direct and exchange) are classical terms.}
    \label{fig:spinless-interaction-channels}
\end{figure}

\subsection{A simple example: Laughlin state and ring-exchange terms}

\begin{figure}[t]
    \centering
    \includegraphics[width=0.75\linewidth]{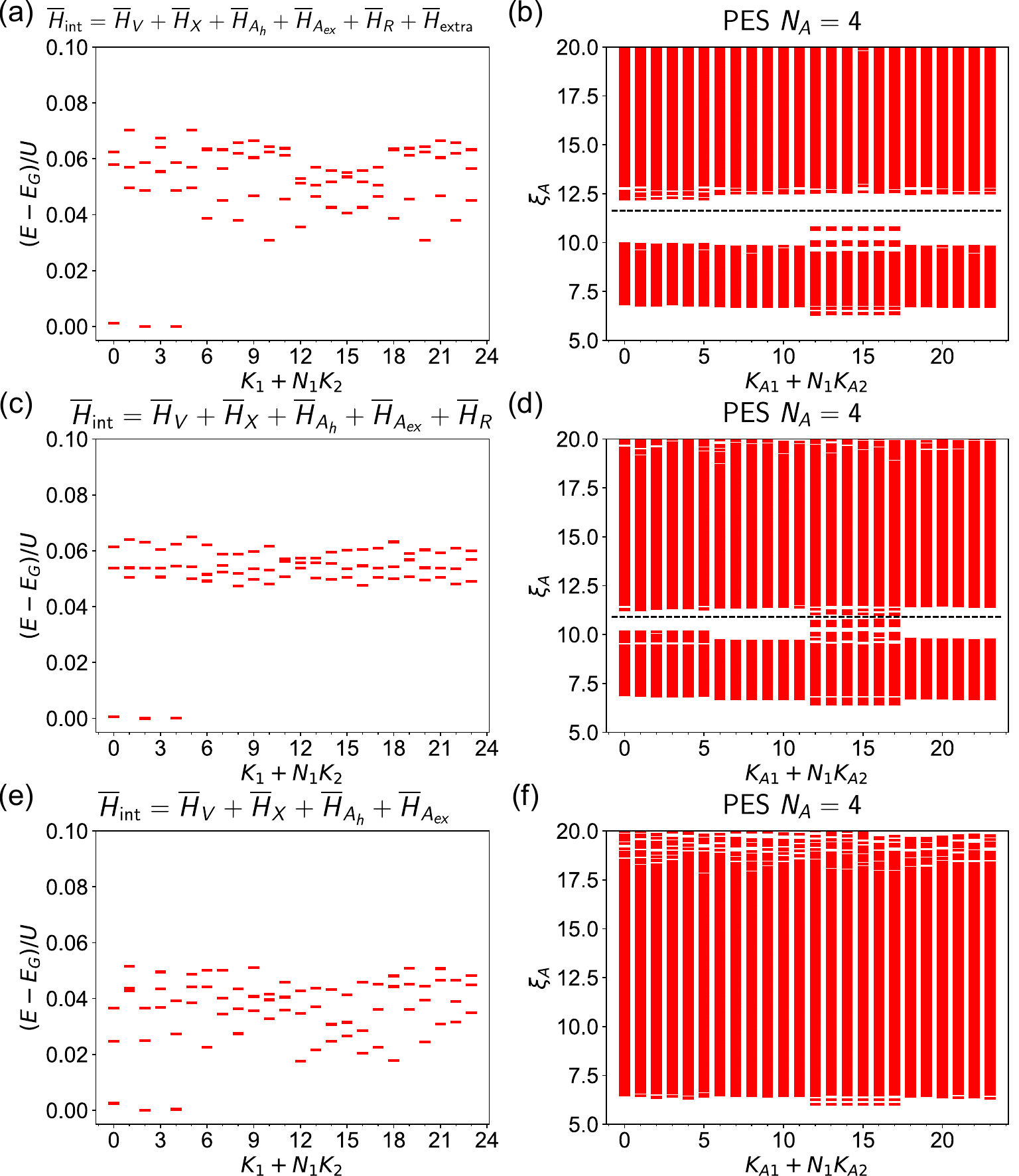}
    \caption{(a) The low energy spectrum of the interacting kagome lattice model with $N=8$ electron solved on a $N_1\times N_2 = 6\times 4$ unit cells using exact diagonalization. Three nearly degenerate eigenstates can be seen at $(K_1, K_2) = (0,0), (2,0)$ and $(4,0)$.
    (b) The particle entanglement spectrum (PES) of the three-fold ground states manifold partitioned into $N_A + N_B = 4 + 4$ electrons. There are $2730$ entanglement spectra eigenvalues below the dashed line, satisfying the $(1, 3)$ admissible partitions-generalized Pauli principle on a torus.
    (c) The energy spectrum of the interacting Hamiltonian in the kagome Chern band with $\overline{H}_{\rm extra}$ removed. 
    (d) The PES of the ground state manifold with $\overline{H}_{\rm extra}$ removed. The amount of entanglement spectrum eigenvalues below the dashed line still satisfy the quasi-hole counting.
    (e) The energy spectrum of the interacting Hamiltonian in the kagome Chern band with both $\overline{H}_R$ and $\overline{H}_{\rm extra}$ removed. 
    (f) The PES of the ground state manifold with both $\overline{H}_R$ and $\overline{H}_{\rm extra}$ removed. There is no longer an entanglement gap in the PES.
    In these simulations, the hopping parameters are chosen as $t = 1$ and $t' = 1/3$. The strength of the nearest neighbor repulsive interaction is chosen as $U = 20t$.
    }
    \label{fig:laughlin-spectrum}
\end{figure}

One of the most astonishing phenomena that one can expect in a nearly flat Chern band is fractional Chern insulator.
The simplest example is the Laughlin wave function, which describes the gapped topological ordered phase of a partially filled lowest Landau level.
In certain context, the bosonic Laughlin states can be interpreted as a Kalmeyer-Laughlin-type chiral spin liquid \cite{Kalmeyer1987, Schroeter2007Spin, Nielsen2012Laughlin}.
As noted in Ref.~\cite{Burkov2010fractional}, the bosonic Laughlin state can be understood as a ground state of a projected interacting Hamiltonian with ring-exchange terms. 
This effective spin model is based on the magnetic Wannier states of the lowest Landau level \cite{Rashba1997Orthogonal}. 
As we have shown in Sec.~\ref{sec:ideal-chern-band}, such magnetic Wannier states already satisfies the condition for optimal gauge choice. Using the algorithm described in Sec.~\ref{sec:gauge}, one may generalize this idea to the interacting spinless fermions projected into a generic Chern band, as opposed to bosons in the lowest Landau level.

Here we use the kagome lattice model, which is discussed in the main text and in Sec.~\ref{sec:kagome}, as an example. 
In Fig.~\ref{fig:laughlin-spectrum}(a), we show the low energy spectrum, which is solved by exact diagonalization, of the fully projected interacting Hamiltonian on a $6\times 4$ unit cell kagome lattice model with $N=8$ electrons. Particle entanglement spectrum (PES) of the three-fold ground states manifold with the partition $N_A + N_B = 4 + 4$ is shown in Fig.~\ref{fig:laughlin-spectrum}(b). The number of entanglement spectrum eigenvalues below the dashed line is $\#{\rm qh} = (N_1N_2)\frac{(N_1N_2-2N_A-1)!}{N_A!(N_1N_2-3N_A)!} = 2730$, satisfying the $(1, 3)$ admissible partitions-generalized Pauli principle on a torus. This ensures that the ground states are indeed Laughlin states.

Using the Wannier functions obtained from a gauge choice with vortex at $\vk_v = (0, 0)$, we can also selectively truncate the terms in the projected interacting Hamiltonian. By removing the generic long-distance four-center interactions $\overline{H}_{\rm extra}$ (excluding the ring-exchange terms $\overline{H}_R$), the ground states manifold, which is shown in Fig.~\ref{fig:laughlin-spectrum}(c), still contains three nearly degenerate states. 
The PES of the ground states manifold with $\overline{H}_{\rm extra}$ removed can also be found in Fig.~\ref{fig:laughlin-spectrum}(d). The number of entanglement spectrum eigenvalues below the dashed line is still $\#{\rm qh} = 2730$, which satisfies the quasi-hole counting. 
However, if the four-center terms, including $\overline{H}_R$, are completely removed, the Laughlin ground states manifold will be destructed, as shown in Fig.~\ref{fig:laughlin-spectrum}(e). The collapsing of the PES gap, shown in Fig.~\ref{fig:laughlin-spectrum}(f), also suggests that the Laughlin states are no longer the ground state of the system.
This indicates that the ring-exchange terms are important for the stability of the Laughlin state in the kagome lattice model, implying that the findings discussed in Ref.~\cite{Burkov2010fractional} could also be pertinent to fermionic systems.

\section{Weierstrass functions}\label{appsec:Elliptic-functions}

\subsection{Definitions}\label{appsec:def}

As we have explained in Sec.~\ref{sec:vortex-gauge}, the Weierstrass $\sigma$ function is necessary when constructing the smooth gauge of Chern band Bloch states. 
In this section, we review the properties of different types of Weierstrass functions that are used in this manuscript. We will use $z$ to represent the complex variable.

By definition, elliptic functions are double-periodic complex functions. We assume the two periods are given by two complex numbers $\omega_1$ and $\omega_2$ that are linear independent. One of the simplest periodic functions that one can write down is:
\begin{equation}
    \wp(z; \mathcal{G}) = \frac{1}{z^2} + \sum_{G\in\mathcal{G}\backslash\{0\}}\left(\frac{1}{(z - G)^2} - \frac{1}{G^2}\right)\,,
\end{equation}
in which the set $\mathcal{G} = \{G| G = m \omega_1 + n\omega_2;m,n\in \mathbb{Z}\}$. Within each unit cell, there is a second order pole, and hence the total residue in the unit cell is zero, which agrees with Liouville's theorem, as discussed in \S20.1 of Ref.~\cite{whittaker1920course}.

The Weierstrass $\zeta$ function is related with the $\wp$ function through the following equation:
\begin{equation}\label{eqn:def-zeta-diff}
    \frac{d\zeta(z; \mathcal{G})}{dz} = -\wp(z; \mathcal{G})\,.
\end{equation}
By performing the integration over the $\wp$ function, one can obtain the definition of $\zeta$ function as follows:
\begin{equation}\label{eqn:def-zeta}
    \zeta(z; \mathcal{G}) = \frac{1}{z} + \sum_{G \in \mathcal{G}\backslash\{0\}}\left(\frac{1}{z - G} + \frac{1}{G} + \frac{z}{G^2}\right)\,.
\end{equation}
Here the $1/G$ term in the summation ensures the {\it absolute convergence} of this infinite series.
Unlike the $\wp$ function, the $\zeta$ function has first order poles sitting at the lattice sites. As a result, $\zeta$ function is no longer a double-periodic function. Instead, it is quasi-double-periodic.

The Weierstrass $\sigma$ function is related to the Weierstrass $\zeta$ function via the following conditions:
\begin{align}
    &\frac{d\log \sigma(z; \mathcal{G})}{d z} = \zeta(z; \mathcal{G})\,,\\
    &\lim_{z\rightarrow 0} \frac{\sigma(z;\mathcal{G})}{z} = 1\,.
\end{align}
By integrating on both sides of Eq.~(\ref{eqn:def-zeta}), we obtain the definition of the Weierstrass $\sigma$ function as the following infinite product:
\begin{equation}\label{eqn:def-sigma}
    \sigma(z;\mathcal{G}) = z\prod_{G\in\mathcal{G}\backslash\{0\}}\left(1 - \frac{z}{G}\right)\exp\left(\frac{z}{G} + \frac{z^2}{2G^2}\right)\,.
\end{equation}
$\sigma(z;\mathcal{G})$ is also an odd quasi-double-periodic function. Moreover, $\sigma(z;\mathcal{G})$ is an {\it entire function}, which does not have poles but only zeros in the complex plane.

\subsection{Periodicity}\label{appsec:periodic}

As we have already mentioned, the $\wp$ function is double-periodic, while both $\zeta$ and $\sigma$ functions are quasi-periodic. Here we provide the proofs of these statements.

We first note that the double-periodicity of $\wp$ function is almost obvious, which is the starting point of proving the periodic properties of the other two functions. For $\zeta$ function, we perform an integral on both sizes of Eq.~(\ref{eqn:def-zeta-diff}), which leads to the following identity:
\begin{equation}
    \zeta(z + \omega_i;\mathcal{G}) - \zeta(z;\mathcal{G}) = \int_{z}^{z+\omega_i} dz'\,\wp(z';\mathcal{G})\,.
\end{equation}
Since the $\wp$ function only has second order poles over the complex plane, the integral on the right hand side does not depend on the integration contour. Due to the periodicity of $\wp$ function, it is also independent to the starting point of the contour.
As such, the right hand side of the above equation is a constant, which is usually denoted as $\eta_i$:
\begin{equation}\label{eqn:zeta-periodic}
    \eta_i = \zeta(z + \omega_i;\mathcal{G}) - \zeta(z;\mathcal{G})\,.
\end{equation}
It is obvious that the value of $\eta_i$ can be obtained by evaluating the $\zeta$ function at $z = \omega_i/2$:
\begin{equation}
    \eta_i = 2\zeta\left(\frac{\omega_i}{2};\mathcal{G}\right)\,,
\end{equation}
which can be proved by noticing that $\zeta$ is an odd function. 

The two parameters $\eta_1$ and $\eta_2$ are related to each other. To prove this, we choose a parallelogram integration contour $c$, whose edges are spanned by $\omega_1$ and $\omega_2$. Since $\zeta$ has one first order pole per unit cell, we obtain the following result using residue's theorem:
\begin{align}
    2\pi i =& \oint_c dz \, \zeta(z;\mathcal{G})\nonumber\\
    =& \int_{z_0}^{z_0 + \omega_1}dz\,\zeta(z;\mathcal{G}) - \int_{z_0 + \omega_2}^{z_0+\omega_2 + \omega_1}dz\,\zeta(z;\mathcal{G}) - \int_{z_0}^{z_0+\omega_2}dz\,\zeta(z;\mathcal{G})+ \int_{z_0+\omega_1}^{z_0 + \omega_1 + \omega_2}dz\,\zeta(z;\mathcal{G})\nonumber\\
    =& \int_{z_0}^{z_0+\omega_1}dz\left[\zeta(z;\mathcal{G}) - \zeta(z+\omega_2;\mathcal{G})\right] + \int_{z_0}^{z_0 + \omega_2}dz\left[\zeta(z+\omega_1,\mathcal{G})-\zeta(z;\mathcal{G})\right]\nonumber\\
    =& -\eta_2\int_{z_0}^{z_0 + \omega_1}dz + \eta_1\int_{z_0}^{z_0 + \omega_2}dz\nonumber\\
    =& \omega_2\eta_1 -\omega_1\eta_2\,.
\end{align}

Finally, by integrating over both sizes of Eq.~(\ref{eqn:zeta-periodic}), we obtain the following equation:
\begin{equation}
    \log\frac{\sigma(z+\omega_i;\mathcal{G})}{\sigma(z;\mathcal{G})} = \eta_i z + C\,,
\end{equation}
in which $C$ is the constant of integration. To determine the value of $C$, we set $z = -\omega_i/2$, and the above equation can be reduced to:
\begin{equation}
     e^{-\eta_i\omega_i/2 + C} = \frac{\sigma(\omega_i/2;\mathcal{G})}{\sigma(-\omega_i/2;\mathcal{G})} =-1 \,.
\end{equation}
Here the second equality is derived from the odd parity of the $\sigma$ function.
As a consequence, the constant of integration is given by:
\begin{equation}
    e^C = -e^{\eta_i\omega_i/2}\,.
\end{equation}
Therefore, we conclude that the quasi-periodic property of the $\sigma$ function is:
\begin{equation}
    \frac{\sigma(z+\omega_i;\mathcal{G})}{\sigma(z;\mathcal{G})} = -e^{\eta_i(z + \omega_i/2)}\,.
\end{equation}
This property of $\sigma$ function is essential to the construction of optimal gauge choice for a Chern band, as we have already addressed in Sec.~\ref{sec:vortex-gauge}.

\subsection{Connection with Jacobi \texorpdfstring{$\vartheta$}{theta} function}\label{appsec:jacobi}

The Weierstrass functions can be related with Jacobi theta function as well. As proved in \S21.43 of Ref.~\cite{whittaker1920course}, the $\sigma$ function has the following representation:
\begin{equation}
    \sigma(z;\mathcal{G}) = \frac{\omega_1}{\pi\vartheta_1'}\exp\left(-\frac{\vartheta_1'''\pi^2z^2}{6\vartheta_1'\omega_1^2}\right)\vartheta_1\left(\frac{\pi z}{\omega_1}\Big{|}\frac{\omega_2}{\omega_1}\right)\,,
\end{equation}
in which $\vartheta_1$ is defined via the following infinite series:
\begin{equation}
    \vartheta_1(z|\tau) = \sum_{n\in\mathbb{Z}}i(-1)^ne^{i\pi\tau n^2}e^{iz(2n+1)}\,,
\end{equation}
and $\vartheta_1'$, $\vartheta_1'''$ represent the first and third order derivatives of the $\vartheta_1$ function with respect to $z$ at $z = 0$ and $\tau=\omega_2/\omega_1$. The defining infinite series of $\vartheta_1$ converges much faster than the infinite product in Eq.~(\ref{eqn:def-sigma}), which is more suitable for numerical computation. The two quasi-periodic parameters can also be represented via Jacobi theta functions as follows:
\begin{align}
    \eta_1 &= -\frac{\pi^2 \vartheta_1'''}{3\omega_1 \vartheta_1'}\,,\\
    \eta_2 &= \frac{\omega_2 \eta_1 - 2\pi i}{\omega_1}\,.
\end{align}

\end{document}